\documentclass[nofootinbib,prd,aps,superscriptaddress,twocolumn]{revtex4-2}
\usepackage{amssymb}
\usepackage{amsfonts}
\usepackage{graphicx}

\usepackage{amsfonts}
\usepackage{amsmath}
\usepackage{amssymb}
\usepackage{amsthm}
\usepackage{bm}
\usepackage{braket}
\usepackage{calc}
\usepackage{cancel}
\usepackage{color}
\usepackage{dsfont}
\usepackage[inline, shortlabels]{enumitem}
\usepackage{fancybox}
\usepackage{fancyhdr}
\usepackage{fixmath}
\usepackage{float}
\usepackage{graphicx}
\usepackage{hyperref}
\usepackage{mathrsfs}
\usepackage{mathtools}
\usepackage{natbib}
\usepackage{slashed}
\usepackage{subcaption}
\usepackage{tcolorbox}										% http://ctan.org/pkg/tcolorbox
\usepackage{textcomp}
\usepackage{transparent}
\usepackage{verbatim}
\usepackage{xcolor}
\usepackage{xparse}
\usepackage{tikz}
\usepackage{tikz-feynman,contour}
\newcommand{\PD}{\partial}
\newcommand{\dbar}{d\hspace*{-0.08em}\bar{}\hspace*{0.1em}}
\begin{document}
%-------------------------------------------------------------------------------------------------------------------------------------------------------% 
\title{Fundamental scalar field with zero dimension from anomaly cancellations}
%-------------------------------------------------------------------------------------------------------------------------------------------------------% 
% \classification{11.15.Tk, 11.10.Ef}
% \keywords      {Schwinger-Dyson Equations, Functional Methods}
%-------------------------------------------------------------------------------------------------------------------------------------------------------% 
\author{J. Miller}
\affiliation{Ariel University, Ariel, 40700, Israel}

\author{G.E.Volovik}
\affiliation{Low Temperature Laboratory, Aalto University, P.O. Box 15100, FI-00076 Aalto, Finland}
\affiliation{Landau Institute for Theoretical Physics, acad. Semyonov av., 1a, 142432, Chernogolovka, Russia}
\author{ M.A.Zubkov }
\affiliation{Ariel University, Ariel, 40700, Israel}

\date{\today}
%-------------------------------------------------------------------------------------------------------------------------------------------------------% 
\begin{abstract}\noindent 
In this article a novel mechanism for dynamical electroweak symmetry breaking 
and the ensuing appearance of fermion mass terms in the action is proposed. 
The action contains 
massless fermions of the SM coupled to gravity through a new type of non-minimal coupling to the vielbein field. 
The corresponding coupling constants in our approach become zero-dimension scalar fields. Such scalar fields provide the cancellation 
of the Weyl anomaly  \cite{Boyle:2021jaz}.    
\end{abstract}
%-------------------------------------------------------------------------------------------------------------------------------------------------------% 
\maketitle
%-------------------------------------------------------------------------------------------------------------------------------------------------------% 
\section{Introduction}\noindent
%-------------------------------------------------------------------------------------------------------------------------------------------------------% 
In this article a novel mechanism of dynamical symmetry breaking, thus leading to the appearance of fermion mass terms in the action, is put forward. 
The assumed action 
contains the full set of standard model (SM) fermions coupled to gravity
through a new type of non-minimal coupling to the vielbein field.
In contrast to the fermion-vielbein coupling via a constant term \cite{Alexandrov:2008iy},
we promote this coupling from a constant to being an actual  field in its own right.
Correspondingly additional distinct scalar fields 
are introduced in the theory
that furnish a set of coupling coefficients%terms
between the various different flavours of SM fermions to the vielbein.
Adding scalar fields to the SM action is known to
imply the cancellation of the Weyl anomaly \cite{Boyle:2021jaz}.
We propose an action 
made up from  the SM fermions
coupled to the vielbein and  fundamental scalar fields
that (i) are  the  very coupling coefficients between the SM  fermions and the vielbein,
(ii) lead directly to the appearance of mass terms of the fermions,
and (iii) generate a theory with a vanishing Weyl anomaly.
This stands as a new model of the composite Higgs Boson.
\par 
%-------------------------------------------------------------------------------------------------------------------------------------------------------% 
The Higgs boson itself has a rich history, beginning with the discovery that
the \emph{Higgs mechanism} \cite{Higgs:1964ia,Higgs:1966ev,Kibble:1967sv,Englert:1964et,Guralnik:1964eu,Weinberg:1967kj}
is responsible for the presence of
mass terms 
in the SM,
where the introduction of the Higgs Boson in the action
spontaneously breaks local symmetries thus ascribing 
mass terms to the massive bosons and fermions.
%-------------------------------------------------------------------------------------------------------------------------------------------------------% 
Explicitly, the SM comprises the 
$SU(3)\times SU(2) \times  U(1)$
gauge model of
strong and electroweak interactions
\cite{Glashow:1961tr,Weinberg:1967tq,salam68,Politzer:1973fx,Cahn:1989by,Fritzsch:1973pi}.
Through the Higgs mechanism  the electroweak $ SU(2) \times  U(1)$ group is spontaneously broken
down to the $U(1)$ symmetry of electromagnetism \cite{Nakada:1958zz,Nambu:1960tm,Schwinger:1962tn,Higgs:1964ia,Englert:1964et,Guralnik:1964eu}.  
Couplings
of the elementary Higgs scalar bosons also break  
quark and lepton
chiral-flavor symmetries,  giving rise to hard mass terms for the quarks and leptons.
The Higgs Boson itself 
has been detected as the scalar excitation with a mass found at 125 GeV \cite{CMS:2012bfw,Mariotti:2015psa,CMS:2012qbp,CMS:2012dun,ATLAS:2012yve,Mountricha:2012cja},
consistent with theoretical predictions.\par 
%-------------------------------------------------------------------------------------------------------------------------------------------------------% 
Despite the success of the Higgs mechanism,
the model admits a number of ambiguities \cite{Lane:2002wv}, perhaps the most striking one is 
the triviality of the model with fundamental scalar fields, as well an emerging
hierarchy problem with vastly disparate energy scales in the theory, and more.
This led to the question, if indeed the Higgs is a new fundamental particle of the SM in its own right,
or rather if the Higgs itself is composed of known fundamental particles of the SM?
\par 
%-------------------------------------------------------------------------------------------------------------------------------------------------------% 
Composite Higgs models began with the suggestion by Terazawa et. al. that
due to its large mass, the top quark is a natural candidate 
for being a
constituent
of the composite Higgs boson \cite{Terazawa:1980nck,Terazawa:1976eq}.
In 1989, the original top-quark condensate 
model of Terazawa et. al. was revived by Miransky, Tanabashi, Yamawaki,
and later by Bardeen, Hill and Lindner \cite{Bardeen:1989ds,Marciano:1989xd,Hill:1991at,Hill:2002ap},
who propesed the top-quark condensate as a means of electroweak symmetry breaking.
In this approach the 
Higgs is composite at short distance scales.
Nevertheless this model 
conflicts with experimental observations on a number of counts.
In these 
models  the energy scale of the new dynamics was assumed to be at about $10^{15}$ GeV. 
Consistent with this value,
these models give  naive  predictions for the Higgs mass at  approximately $2 m_t \sim 350$ GeV 
\cite{Terazawa:1976eq,Miransky:1989ds,Bardeen:1989ds,Marciano:1989xd,Hill:1991at,Hill:2002ap}, 
and present experimental data 
rules out such a value.
A smaller value of the Higgs boson mass might be obtained if the calculations  
rely 
on large renormalization group corrections \cite{Bardeen:1989ds,Marciano:1989xd,Hill:1991at,Hill:2002ap},
due to the running of coupling constants between the 
working scale ($10^{15}$ GeV) and the electroweak scale ($100$ GeV). 
At any rate
this running is not able to explain the observed appearance of the Higgs  
mass at around 125 GeV.
Even more these models require fine tuning of numerous parameters in order to match experimental results. 
But that aside, the models with  four - fermion interactions should be taken as the phenomenological ones, 
in which only the leading order in the $1/N_c$ expansion is taken into account. 
\par 
%-------------------------------------------------------------------------------------------------------------------------------------------------------% 
The \emph{top see-saw} model was 
advocated initially by Chivukula, Dobrescu, Georgi and Hill \cite{Chivukula:1998wd} in response to the difficulties of the earlier top-quark condensate models.
The model admits, in addition to the  top quark, an additional heavy fermion, $\chi$ \cite{Dobrescu:1997nm,Chivukula:1998wd}.
In particular 
in the top see-saw model there is no need for fine tuning of parameters,
and a later version of the model by Dobrescu and Cheng \cite{Cheng:2013qwa} includes a light composite Higgs boson.
\par 
%-------------------------------------------------------------------------------------------------------------------------------------------------------% 
In the 1970s the technicolor (TC) model was formulated 
\cite{Susskind:1978ms,Weinberg:1975gm} as an attempt to resolve the shortcomings of the SM already discussed, 
while providing sound theoretical predictions 
for the Higgs mass.
TC  is
a gauge theory of fermions with no elementary scalars,
yet at the same time inclusive of 
dynamical electroweak symmetry breaking  and flavor symmetry breaking.
In the TC model the Higgs, instead of being a fundamental scalar is composed of a new class of fermions called technifermions
interacting via technicolor gauge bosons. This interaction is attractive and as a result, 
by analogy with Bardeen-Cooper-Schrieffer (BCS) superconductor theory, leads to the formation of fermion condensates. 
However, TC theory 
fails to produce mass terms for the quarks and leptons.
For this reason the model was extended to the Extended Technicolor (ETC) model \cite{Dimopoulos:1979es,Eichten:1979ah},
but this model is inconsistent with experimental constraints on flavor changing neutral currents,
and precision electroweak measurements.
Later on an improved TC model called the walking technicolor model
\cite{Appelquist:1998rb,Gudnason:2006ug} 
was constructed
that resolves these issues, although it still fails to predict correctly  the measured value of the top-quark mass.
For a detailed pedagogical review of the TC and ETC models see Refs. \cite{Lane:2002wv,Lane:1993wz}.
\par 
%-------------------------------------------------------------------------------------------------------------------------------------------------------% 
After the technicolor models came  
the idea that the Higgs boson appears as a pseudo-Goldstone boson, as noted originally in Ref. \cite{Dugan:1984hq}. This idea forms the basis of 
\emph{Little Higgs Models}
\cite{Perelstein:2005ka,Arkani-Hamed:2002ikv,Low:2002ws,Berger:2005ht,Redi:2012ha,Arhrib:2020tqk}. 
The model stems a well known result of
Goldstone's theorem, namely that spontaneous breaking of a global symmetry yields massless scalar particles, or \emph{Goldstone bosons}.
By selecting the right global symmetry, it is possible to have Goldstone bosons that correspond to the Higgs doublet in the SM.
\par 
%-------------------------------------------------------------------------------------------------------------------------------------------------------% 
Subsequently the partial compositeness model was conjectured,   by Kaplan
\cite{Kaplan:1991dc} initially. Therein
each SM particle has a heavy partner that can mix with it. 
Like this the SM particles are linear combinations of elementary and composite states
via 
a mixing angle.
The elegance of the partial compositeness model is its simplicity,
without the need for deviations beyond the standard model, thus  relying 
on the known fundamental particles only \cite{Redi:2011zi}.
\par 
%-------------------------------------------------------------------------------------------------------------------------------------------------------% 
More recently, among a variety of theoretical papers that appeared between December 2015 and August 2016,  there are several that consider both
the $125$ GeV Higgs boson, and the hypothetical new heavier Higgs boson as composite due to the new strong interaction, 
as noticed  for example in Refs. 
\cite{Hong:2016uou,Matsuzaki:2016joz}. 
In a different approach there are other papers devoted to the description of the composite nature of the heavier ($750$ GeV) Higgs boson only 
\cite{Harigaya:2016pnu,
Nakai:2015ptz,
Franceschini:2015kwy,
Molinaro:2015cwg,
Bian:2015kjt,
Bai:2015nbs,
Cline:2015msi,
Ko:2016wce,
Harigaya:2016pnu,
Redi:2016kip,Harigaya:2016eol,
Foot:2016llc,
Iwamoto:2016ral,
Bai:2016vca,Kobakhidze:2015ldh,
Foot:2016llc}.
\par 
%-------------------------------------------------------------------------------------------------------------------------------------------------------% 
Other  more modern composite Higgs  models consist of  
the walking model,
\cite{Holdom:1988gs,Holdom:1989aa}
the ideal walking  model
\cite{Yamawaki:1996vr,Fukano:2010yv,Rantaharju:2017eej,Rantaharju:2019nmh},
the technicolor scalar model
\cite{Foadi:2012bb}, 
Sannino's model of the generalized orientafold gauge theory approach to electroweak symmetry breaking
\cite{Sannino:2004qp},
and 
the walking model in higher-dimensional $SU(N)$ gauge theories
of Dietrich and Sannino
\cite{Dietrich:2006cm,Dietrich:2005jn}.
A more comprehensive review of modern composite Higgs models can be found in 
\cite{Cacciapaglia:2020kgq}.
\par 
%-------------------------------------------------------------------------------------------------------------------------------------------------------% 
Quantum gravity in the first order formalism with
either the Palatini action \cite{palatini1919,RovelliPalatini} or the Holst action \cite{Holst:1995pc},
shares a property with the above composite Higgs models: 
it
leads to a four - fermion interaction.
This four fermion interaction 
is now between spinor fields
coupled in a minimal way to the torsion field \cite{Perez:2005pm,Rovelli4fermion}, 
but importantly 
this
four - fermion interaction leads to fermion condensates 
of the type that could constitute the composite Higgs. 
%-------------------------------------------------------------------------------------------------------------------------------------------------------% 
A similar idea has already been 
suggested in Ref. \cite{Zubkov:2010sx}, namely that the torsion field coupled in a non-minimal way
to fermion fields \cite{Belyaev:1998ax,Shapiro:1994vs,Shapiro:2001rz}  could be a source of
dynamical electroweak symmetry breaking.
\par 
%-------------------------------------------------------------------------------------------------------------------------------------------------------%
This type of action with  fermion fields coupled to the vielbein
is one of the main ingredients of our action.
%-------------------------------------------------------------------------------------------------------------------------------------------------------%
The other main ingredient flows from 
a theory that contains 
fundamental scalar fields with zero mass-dimension recently proposed in 
\cite{Boyle:2021jaz},
as a way of simultaneously canceling both the vacuum energy and the Weyl anomaly.
Starting with 
a  theory of the  SM  coupled to gravity,
the fermion and gauge fields  are coupled to a background classical gravitational 
field, with
a right-handed neutrino for each generation also coupled to the background field.
This representation is a QFT on a classical background spacetime.
The Weyl anomaly
is a measure of the failure of the classically Weyl-invariant theory
to define a Weyl-invariant quantum theory,
and can be expressed in terms of the number of different fields present in the theory.
In the SM there are $n_{0}=4$ ordinary real scalars in the usual complex Higgs doublet, $n_{1/2}=3\times16=48$ Weyl spinors (16 per generation), $n_{1}=8+3+1=12$  gauge fields of
$SU(3)\times SU(2)\times U(1)$, and a single 
gravitational field ($n_{2}=1$).
With these combinations the Weyl anomaly is non-zero.
However suppose that the Higgs and graviton fields are not fundamental fields but rather composite, such that their contributions to the vacuum energy and Weyl anomaly 
can be dropped, and  $n_{0}=n_{2}=0$.  The implication is that by introducing $n_{0}'=36$  scalars with zero mass-dimension, 
then not only does the  Weyl anomaly vanish, but also   
the vacuum energy vanishes as well.  
\par 
%-------------------------------------------------------------------------------------------------------------------------------------------------------% 
This embodies our motivation for the starting point of 
this work: a model comprising
Dirac fermions of the SM 
with a non-minimal coupling to
gravity, that includes 
36 fundamental scalar fields.
The details of the construction of the theory is
described in  full in 
\S\ref{sec_nonminimal_coupling_of_fermions_to_gravity}.
We find that in this model,
the theory admits  mass terms for the fermions.
By using the Schwinger-Dyson equation for the fermion self-energy function,
a relationship is derived between the top-quark mass, the Planck mass (the ultraviolet cut-off point of the loop integral) and the strength of the inverse coupling between the fermion fields and the vielbein.
From this relation 
an estimation for the mass of the top-quark can be extracted.\par 
%-------------------------------------------------------------------------------------------------------------------------------------------------------% 
This paper is organized as follows.
In Section \ref{sec_fundamental_scalars_in_models_of_3_generations_of_SM_fermions}
the first main ingredient of the action is constructed, the piece that includes Dirac fermions coupled non-minimally to the vielbein field
that describes the background gravitational field.
In \S\ref{sec_EW_symmetry_breaking_and_mass_generation} the remaining piece of
the action is formulated, namely the dynamical piece of the fundamental scalar fields described above.
In \S\ref{sec_SD_for_calculating_mf}
the Schwinger-Dyson approach is implemented to 
calculate the leading-order piece of the self-energy of the dominant scalar field in the action,
namely the scalar field that couples to the top-quark, being the heaviest flavour.
In this section the main result of the paper is derived, namely 
the emergence of a mass term for the top quark.
A relationship between the top-quark mass, the coupling of the 
scalars to the vielbein (gravitational) field, and the cut-off energy of the loop integral (the Planck mass)
is used to estimate the leading-order contribution to the top-quark mass.
In \S\ref{sec_conclusion} the main conclusions of the paper are summarized along with prospects for future research.
There are two appendices: appendix \S\ref{sec_wick_rotatns} contains the main formulas for a Wick rotation used in the computation of the one-loop integral,
and appendix \S\ref{sec_angular_integral} contains full details of the angular integrals in the one-loop integral.
\par
%-------------------------------------------------------------------------------------------------------------------------------------------------------% 
Throughout this article
lower case latin letters
$a,b,c\dots=0,1,2,3$
label internal Lorentz indices,
greek letters 
$\mu,\nu,\dots=0,1,2,3$
label spacetime indices, and 
$\epsilon_{abcd}$ is the completely antisymmetric Levi-Civita symbol.
Lorentz indices are raised and lowered by the Minkowski metric $\eta_{ab}$ and 
spacetime indices are raised and lowered by a spacetime metric $g_{\mu\nu}$.
Metrics are assumed to have a mostly negative signature, 
and specifically the Minkowski metric is
$\eta_{ab}={\rm diag}(1,-1,-1,-1)$.
%-------------------------------------------------------------------------------------------------------------------------------------------------------% 
\section{Fundamental scalars in models with three generations of SM fermions}
\label{sec_fundamental_scalars_in_models_of_3_generations_of_SM_fermions}\noindent
%-------------------------------------------------------------------------------------------------------------------------------------------------------% 
\subsection{Non-minimal coupling of fermions with gravity}
\label{sec_nonminimal_coupling_of_fermions_to_gravity}\noindent
%-------------------------------------------------------------------------------------------------------------------------------------------------------% 
The action for Dirac fermions coupled to the vielbein field can be written as
\cite{Rovelli:2014ssa,Perez:2005pm}
%-------------------------------------------------------------------------------------------------------------------------------------------------------% 
\begin{equation}S=\frac1{6}\epsilon_{abcd}\int d^4x\, {{{\theta}}}^a\wedge e^b\wedge e^c\wedge e^d\ ,\label{actn_dirac_ferms_1}\end{equation}
%-------------------------------------------------------------------------------------------------------------------------------------------------------% 
where
%-------------------------------------------------------------------------------------------------------------------------------------------------------% 
\begin{equation} {{{\theta}}}^a=\frac i{2}\left[\bar\psi\gamma^aD_\mu\psi-
\overline{D_\mu \psi}\gamma^a\psi\right] dx^\mu \ ,\label{actn_dirac_ferms_2}\end{equation}
%-------------------------------------------------------------------------------------------------------------------------------------------------------% 
with $\gamma^a$ the usual Dirac matrices,
$\bar\psi=\psi^\dagger\gamma^0$, and $D_\mu\psi$ is the covariant derivative acting on 
the Dirac fermion field defined by \cite{Rovelli04}
%-------------------------------------------------------------------------------------------------------------------------------------------------------% 
\begin{equation} D_\mu\psi=\PD_\mu\psi-\frac i{2}\gamma_{ab}\omega^{ab}_{\ \ \mu}\psi\ 
\ ,\label{cov_der}\end{equation}
%-------------------------------------------------------------------------------------------------------------------------------------------------------% 
where $\gamma_{ab}=\frac1{4}[\gamma_a,\gamma_b]$ and $\omega^{ab}_{\ \ \mu}$ is the spin connection.
The fields $e^a$ are the vielbein fields that describe the background gravitational field,
related to the spacetime metric $g$ through the completion relation
$\eta_{ab}e^a_{\ \mu}e^b_{\ \nu}=g_{\mu\nu}$.
The action in (\ref{actn_dirac_ferms_1}) is %in first-order form in the sense that it is 
a functional of $e$ alone, as opposed to being a functional of both $e$ and $\omega$.
Explicitly, 
%in the first-order formalism 
$\omega$ becomes dependent on $e$ 
by imposing  
Cartan’s first structure equation.
%with a fermion source on the right, which comes from varying the connection in the action.
\par
%-------------------------------------------------------------------------------------------------------------------------------------------------------% 
In turn, the action for the right - handed Weyl fermions
may be written as Eq. (\ref{actn_dirac_ferms_1}) with ${{{\theta}}}^a \to  {{{\theta}}} ^a_R$, where
%-------------------------------------------------------------------------------------------------------------------------------------------------------% 
\begin{equation} 
{{{\theta}}}^a_R=\frac i{2}\left[\bar\psi_R\sigma^aD_\mu\psi_R-\overline{D_\mu \psi}_R\sigma^a\psi_R\right] dx^\mu \ ,\label{actn_dirac_ferms_3}\end{equation}
%-------------------------------------------------------------------------------------------------------------------------------------------------------% 
while the action for the left - handed Weyl fermions may
be written as   Eq. (\ref{actn_dirac_ferms_1}) with ${{{\theta}}}^a \to  {{{\theta}}} ^a_L$ and
%-------------------------------------------------------------------------------------------------------------------------------------------------------% 
\begin{equation} {{{\theta}}}^a_L=\frac i{2}\left[\bar\psi_L\bar \sigma^aD_\mu\psi_L-\overline{D_\mu \psi}_L
\bar \sigma^a\psi_L\right] dx^\mu \ ,\label{actn_dirac_ferms_3}\end{equation}
%-------------------------------------------------------------------------------------------------------------------------------------------------------% 
A non - minimal coupling of fermions to gravity 
is achieved 
via the substitution
\cite{Alexandrov:2008iy}
%-------------------------------------------------------------------------------------------------------------------------------------------------------% 
\begin{equation} {{{\theta}}}^a_R\to
{{{\theta}}}^a_R
+\frac i{2}\left[ \bar\psi_R \xi_R\sigma^aD_\mu\psi_R
-
\xi^\ast_R\overline{ D_\mu \psi}_R \sigma^a\psi_R\right] dx^\mu 
\label{non_min_cplng_1}\end{equation}
%-------------------------------------------------------------------------------------------------------------------------------------------------------% 
and 
%-------------------------------------------------------------------------------------------------------------------------------------------------------% 
\begin{equation} {{{\theta}}}^a_L\to
{{{\theta}}}^a_L
+\frac i{2}\left[ \bar\psi_L \xi_L\bar \sigma^aD_\mu\psi_L
-
\xi^\ast_L \overline{D_\mu \psi}_L \bar \sigma^a\psi_L\right] dx^\mu  \label{non_min_cplng_2}\end{equation}
%-------------------------------------------------------------------------------------------------------------------------------------------------------% 
with  complex - valued constants $\xi_R$ and $\xi_L$. 
Notice that in the absence of the spin connection, $D_\mu=\PD_\mu$ and the imaginary
parts of  $\xi_R$ and $\xi_L$ decouple and do not interact with
fermions. 
At the same time the real parts of these constants may be absorbed by a rescaling of the fermion fields.
\par 
%-------------------------------------------------------------------------------------------------------------------------------------------------------% 
That said, suppose that this construction is extended to the
case where  $\xi_R$ and $\xi_L$ become coordinate-dependent
fields. In this case
both the real
and imaginary parts of $\xi_R$ and $\xi_L$  interact
with the fermions of the theory,
and cannot be removed by any rescaling of the
fields. 
Further more,  let $\xi_R$ and $\xi_L$ 
be assigned flavour indices.
In the presence of $SU (3)\otimes SU (2) \otimes  U (1)$
gauge fields, there are 
$n_1 = 8 + 3 + 1 = 12$
vector fields. In accordance the number of Weyl fermions must equal
$
n_{1/2} = 4n_1 = 48$.
This number coincides with the number of Weyl fermions
in the SM with three generations: 
$n_{ 1/2} = (1_{\rm leptons} +
3_{\rm quarks} ) \times  2_{{\rm up}\,{\rm \&}\, {\rm down}} \times  2_{{\rm left}\, {\rm\&}\,{\rm right}} \times  3_{\rm generations} = 48$.
%-------------------------------------------------------------------------------------------------------------------------------------------------------% 
\subsection{Parity breaking interactions with zero dimension scalar fields}
\label{sec_parity_breaking_interactions_zero_dim_scalar_fields}\noindent
%-------------------------------------------------------------------------------------------------------------------------------------------------------% 
The theory can be extended one stage further to the case where 
$\xi_L$ and $\xi_R$ carry not only flavor indices but also 
generation indices.
Even more they can be further distinguished by assigning  
different forms for the matrices $\xi$
for the left - handed and the right - handed particles, while
$\xi$ the matrices for the quarks and leptons remain identical. 
In this approach the fermion term $\theta^a$ in (\ref{actn_dirac_ferms_1}) now reads
\begin{equation}{{{\theta}}}^a= {{{\theta}}}^a_L+ {{{\theta}}}^a_R \ ,\label{parity_brkng_zero-dim_sclr_flds}\end{equation}
%-------------------------------------------------------------------------------------------------------------------------------------------------------% 
where 
%-------------------------------------------------------------------------------------------------------------------------------------------------------% 
\begin{align} {{{\theta}}}^a_R=&
\frac i{2} \bigg[\bar\psi_R \left(1+\xi_R\right)\sigma^aD_\mu\psi_R
\nonumber\\&\qquad\qquad- (D_\mu\bar\psi_R)\left(1+
\xi^\dagger_R\right)\sigma^a\psi_R \bigg] dx^\mu \ ,\label{parity_brkng_zero-dim_sclr_flds_1}\end{align}
%-------------------------------------------------------------------------------------------------------------------------------------------------------% 
and 
%-------------------------------------------------------------------------------------------------------------------------------------------------------% 
\begin{align} {{{\theta}}}^a_L =&
\frac i{2}\bigg[ \bar\psi_L \left(1+
\xi_L\right)\bar \sigma^aD_\mu\psi_L
\nonumber\\ &\qquad\qquad-
(D_\mu\bar\psi_L)\left(1+
\xi_L^\dagger\right)\bar \sigma^a\psi_L\bigg]
dx^\mu \ .\label{parity_brkng_zero-dim_sclr_flds_2}\end{align}
%-------------------------------------------------------------------------------------------------------------------------------------------------------% 
This gives rise to
$n_0'  = 2_{\rm chiralities} \times  2_{{\rm real}\,{\rm \&}\,{\rm imaginary}\,{\rm parts}} \times 
3_{\rm generations}  \times  3_{\rm generations}  = 36$
components of the scalar
fields. This is precisely the number of scalars $n_0'=3n_1$ 
needed for the cancellation of the Weyl anomaly
\cite{Boyle:2021jaz}.
\par 
%-------------------------------------------------------------------------------------------------------------------------------------------------------% 
\subsection{Interactions that conserve parity}\label{sec_interactions_that_conserve_parity}\noindent
%-------------------------------------------------------------------------------------------------------------------------------------------------------% 
There would be the same number of scalar fields as
above if $\xi_L$ were identical to $\xi_R$, but with the matrices $\xi$ being
different for quarks and leptons. In this case the interaction
with $\xi$ does not break parity. The fermion
action is still defined as in  Eq. (\ref{actn_dirac_ferms_1}), but instead of Eq. (\ref{parity_brkng_zero-dim_sclr_flds}), 
%-------------------------------------------------------------------------------------------------------------------------------------------------------% 
\begin{equation}{{{\theta}}}^a= {{{\theta}}}^a_l+ {{{\theta}}}^a_q \ ,\label{10}\end{equation}
%-------------------------------------------------------------------------------------------------------------------------------------------------------% 
where 
%-------------------------------------------------------------------------------------------------------------------------------------------------------% 
\begin{align} {{{\theta}}}^a_q=&
\frac i{2}\bigg[ \bar\psi_q \left(1+
\xi_q\right)\gamma^aD_\mu\psi_q
\nonumber\\&\qquad\qquad 
-
(D_\mu\bar\psi_q)\left(1+
\xi_q^\dagger\right)\gamma^a\psi_q\bigg] dx^\mu \ ,\label{11}\end{align}
%-------------------------------------------------------------------------------------------------------------------------------------------------------% 
and 
%-------------------------------------------------------------------------------------------------------------------------------------------------------% 
\begin{align} {{{\theta}}}^a_l=&
\frac i{2}\bigg[ \bar\psi_q \left(1+
\xi_l\right)\gamma^aD_\mu\psi_l
\nonumber\\&\qquad\qquad
-
(D_\mu\bar\psi_l)\left(1+
\xi_l^\dagger\right)\gamma^a\psi_l\bigg] dx^\mu \ .
\label{parity_conservatn_zero-dim_sclr_flds_2}\end{align}
%-------------------------------------------------------------------------------------------------------------------------------------------------------% 
Here both $\xi_l$ and $\xi_q$ are the $3 \times  3$ complex - valued
matrices in flavor space. As before, 
$n_0'  = 2_{{\rm leps}\,{\rm \&}\,{\rm qrks}} \times  2_{{\rm real}\,{\rm \&}\,{\rm imag}\,{\rm parts}} \times 
3_{\rm generations}  \times  3_{\rm generations}  = 36$
zero dimension scalar fields.\par 
%-------------------------------------------------------------------------------------------------------------------------------------------------------% 
\subsection{$\Phi$ - fields}
%-------------------------------------------------------------------------------------------------------------------------------------------------------% 
It is worth mentioning that $\theta$ - field of Eq. (\ref{11}) may be replaced in a different way in order to produce the zero dimension scalar fields:
\begin{eqnarray}
\theta^a \to \frac{i}{2} \left[ \bar\psi \gamma^a D_\mu \psi - (D_\mu \bar\psi)\gamma^a \psi\right] dx^\mu +
\nonumber
\\
+
\frac{i}{2}   \left[ \bar\psi_L^A \Phi_{AB}^a D_\mu \psi^B_R - (D_\mu \bar\psi_R)^B  \Phi_{AB}^{+,a} \psi_L^A \right] dx^\mu
\,.
\label{Action3}
\end{eqnarray}
Here $\Phi$ is the $SU(2)$ doublet that carries in addition the generation index $A$ and the flavor $B$ of the right - handed fermions. (We may take $e,\mu, \tau$ for leptons, and $d,s,b$ and/or $u,c,t$ for quarks.) This expression is the further extension of the non - minimal coupling of fermions to gravity, in which the above mentioned constants $\xi$ not only become fields, but are, in addition, transformed under the $SU(2)_L$ group. For example, we can take $A = 1,2,3$ corresponding to the three generations of quarks, and $B = 1,2,3$ corresponding to the $u,c,t$ quarks. As a result we  obtain  $4\times 4\time 3 \times 3 = 144$ real - valued components of these scalar fields. This number is larger than the one needed for the cancellation of Weyl anomaly \cite{Boyle:2021jaz}. The minimal choice here is if we consider the singlets in flavor space, which gives only 
$4 \time 4 = 16$ components of the scalars - the number that is smaller than the required $32$ components. This scheme, therefore cannot be used directly to build the theory with the cancellation of Weyl anomaly. The other ingredients should be added in  order to match this condition. Therefore, in the following we will concentrate on the scheme proposed above in Sect.  \ref{sec_interactions_that_conserve_parity}.\par
%-------------------------------------------------------------------------------------------------------------------------------------------------------% 
\section{Electroweak symmetry breaking and mass generation}
\label{sec_EW_symmetry_breaking_and_mass_generation}
\noindent
%-------------------------------------------------------------------------------------------------------------------------------------------------------% 
Consider the SM with three generations and the
fundamental scalar $\xi$ fields constructed in \S \ref{sec_interactions_that_conserve_parity}.
With variations of the vielbein and   spin connection
restrained,  define an action for the zero dimension scalar fields
as
%-------------------------------------------------------------------------------------------------------------------------------------------------------% 
\begin{equation}S_B=\alpha^{uv}_{ABCD}\int d^4x\, 
\left(\xi_u^{AB}\right)^\ast \Box^2 \xi^{CD}_v\ ,\label{S_B}\end{equation}
%-------------------------------------------------------------------------------------------------------------------------------------------------------% 
where $A,B,C,D=1,2,3$ are generation indices,  $u,v = q, l$ are flavor indices, and $\alpha^{uv}_{ABCD}$
are a set of coupling constants.
%-------------------------------------------------------------------------------------------------------------------------------------------------------% 
The effective four - fermion interactions arise from the 
exchange by quanta of the fields $\xi_q$ and $\xi_l$.  
The effective four - fermion interaction is non-local.\par
\begin{figure}[htb!]
\begin{center} 
\begin{tikzpicture}
\begin{feynman}
\vertex (i1) at (0,0) ;
\vertex (v1) at (2,0) ;
\vertex (v2) at (4,0) ;
\vertex (v1a) at (2,0.01) ;
\vertex (v2a) at (4,0.01) ;
\vertex (v1b) at (2,0.02) ;
\vertex (v2b) at (4,0.02) ;
\vertex (f1) at (6,0) ;

\diagram* {
(i1) --  [fermion, edge label=\(p\)](v1),
(v2) --  [fermion,edge label=\(p\)](f1),
(v1) -- [edge label=\(k\)](v2),
(v1a)-- (v2a),
(v1b)-- (v2b),
(v1) -- [scalar, half left,edge label=\(p-k\)] (v2)
};
\end{feynman}
\end{tikzpicture}\end{center}
\caption{\footnotesize 
Feynman diagram of the fermion self-energy corresponding to Eq.~(\ref{Sigma_p}).
The incoming and outgoing lines are initial and final fermion states with momentum $p$.
The dashed line in the loop is a scalar boson propagator carrying virtual momentum $p-k$, and the thick
horizontal line corresponds to the full fermion propagator with renormalized mass $\Sigma (k)$
inclusive of 
all self-energy corrections, as related in Eq.~(\ref{Sigma_p}).
}\label{fig_1}
\end{figure}
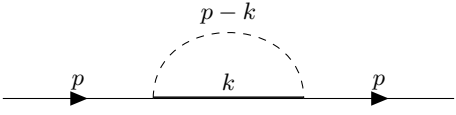

%-------------------------------------------------------------------------------------------------------------------------------------------------------% 
Consider now a certain sector of the theory that describes 
the dominant contributions of the field $\xi_q^{33}$.
Add this piece of the action to the action in (\ref{actn_dirac_ferms_1}),
with $\theta^a$ given by (\ref{10}), to obtain
%-------------------------------------------------------------------------------------------------------------------------------------------------------% 
\begin{align}S=&\int d^4x\,\bigg[\alpha \xi^\ast \Box^2 \xi+\frac i{2}\bar Q(1+\xi)\slashed D Q
\nonumber\\&\qquad\qquad-\frac i{2}\overline{\slashed D Q} (1+\xi^\ast)Q 
\bigg]
\end{align}
%-------------------------------------------------------------------------------------------------------------------------------------------------------% 
where $\slashed D=\gamma^\mu D_\mu$, $\xi=\xi_q^{33}$, $\alpha=\alpha^q_{3333}$ and
%-------------------------------------------------------------------------------------------------------------------------------------------------------% 
\begin{equation}Q_3(x)=
\begin{pmatrix}t\\b\end{pmatrix}
\ \label{Q3}\end{equation}
%-------------------------------------------------------------------------------------------------------------------------------------------------------% 
is the third generation of quarks. Note that the covariant
derivative $D$ contains gauge fields.
Define an effective action, $S_4$
as
$e^{iS_4}=\frac1{Z_0}\int D\xi\,D\xi^\ast e^{iS}$.
After integrating out the scalar fields 
the  corresponding effective action is found to be 
%-------------------------------------------------------------------------------------------------------------------------------------------------------% 
\begin{align}S_4=&-\frac1{4\alpha}\int d^4x\,d^4y\,
\bar Q_3(x)\gamma^\mu D_\mu Q_3(x)
\nonumber\\
&
\Box^{-2}(x,y)D_\nu \bar Q_3(y)\gamma^\nu Q_3(y)
\  ,\label{S_4}\end{align}
%-------------------------------------------------------------------------------------------------------------------------------------------------------% 
where $\Box^{-2}(x,y)$ is the square of the inverse propagator of the boson field $\xi^{33}_q$.
The one-loop contribution to the two-point Green function is straightforwardly read off Eq.~(\ref{S_4}).
The form of the self - energy of the third generation quarks
then follows, as written in  
Eq.~(\ref{Sigma_p}) below.
The corresponding Feynman diagram is shown in Fig.~\ref{fig_1}.
%-------------------------------------------------------------------------------------------------------------------------------------------------------%
\par 
%-------------------------------------------------------------------------------------------------------------------------------------------------------%
\section{Schwinger-Dyson approach for calculating fermion mass}
\label{sec_SD_for_calculating_mf}
%-------------------------------------------------------------------------------------------------------------------------------------------------------%
At this point the discussion follows the method of
\cite{Zubkov:2014qla}, itself based on \cite{Miransky_1}.
The idea of this method is to 
write down
the Schwinger-Dyson equation in order express the inverse of the full propagator, $D^{-1}(p)$
in terms of the self-energy function $\Sigma(p)$ 
in the form of a simplified ansatz, through which  $\Sigma(p)$ can be determined.
The self - energy function,
$\Sigma (p)$ of the third generation quarks through leading order,
has the form
%-------------------------------------------------------------------------------------------------------------------------------------------------------% 
\begin{equation}
\Sigma(p)=\frac 1{\alpha }\int \dbar^4k\,\gamma k \frac{i  }{\gamma k- \Sigma(k)}\gamma k
\frac1{(p-k)^4}\ ,
\ \label{Sigma_p}\end{equation}
%-------------------------------------------------------------------------------------------------------------------------------------------------------% 
where 
the standard notation
$\int\dbar^4k\left(\dots\right)=(2\pi)^{-4}\int d^4k\left(\dots\right)$ has been  used.
Now apply a Wick rotation (see Appendix \ref{sec_wick_rotatns} for details) to obtain 
%-------------------------------------------------------------------------------------------------------------------------------------------------------% 
\begin{align}
i\Sigma(p)
=&\frac {1}{\alpha }\int \dbar^4k_E\, \gamma_Ek_E \frac{1}{\left(\gamma_E k_E-i\Sigma(k)\right)}\gamma_Ek_E 
\nonumber\\&\times\frac1{(p_E-k_E)^4}
\ .\label{Sigma_p_wick3A}\end{align}
%-------------------------------------------------------------------------------------------------------------------------------------------------------%
Since all expressions from now on are in terms of Euclidean-space variables 
the subscript `$E$' can be  dropped.\par
%-------------------------------------------------------------------------------------------------------------------------------------------------------% 
The ansatz for the Schwinger-Dyson equation has the form
%-------------------------------------------------------------------------------------------------------------------------------------------------------%
\begin{equation}
D^{-1}(p)=A(p^2)\gamma  p  -i B(p^2)=\gamma  p  -i \Sigma (p)\ .
\label{SD_eqn}
\end{equation}
%-------------------------------------------------------------------------------------------------------------------------------------------------------%
The right-hand side of (\ref{SD_eqn}) may be substituted inside (\ref{Sigma_p_wick3A}). 
To keep notation brief it is useful to use the shorthand 
$A=A(k^2)$ and $B=B(k^2)$.
The resulting expression is
%-------------------------------------------------------------------------------------------------------------------------------------------------------% 
\begin{align}
i\Sigma(p)
%-------------------------------------------------------------------------------------------------------------------------------------------------------%  
=&\frac {1}{\alpha }\int \dbar^4k\,\gamma k
\frac{\left( A{} \gamma  k  +i  B{} \right) }{\left(A^2{} k^2+ B^2{}\right) }  \gamma k \frac1{ (p-k)^4}\ .
\label{SD_eqn_A}
\end{align}
%-------------------------------------------------------------------------------------------------------------------------------------------------------% 
By (\ref{wick_rot_3A}), 
$(\gamma k)^2=\gamma^\mu\gamma^\nu k_\mu k_\nu=\frac1{2}\delta^{\mu\nu}k_\mu k_\nu=k^2$, and hence (\ref{SD_eqn_A})
becomes 
%-------------------------------------------------------------------------------------------------------------------------------------------------------% 
\begin{align}
i\Sigma(p)
=&\Sigma_1(p)+\Sigma_2(p)\ ,\label{SD_eqn_2b}\end{align}
%-------------------------------------------------------------------------------------------------------------------------------------------------------%
where 
%-------------------------------------------------------------------------------------------------------------------------------------------------------% 
\begin{align}
\Sigma_1(p)
:=&\frac { 1}{\alpha }\int \dbar^4k\,\gamma k\,
\frac{k^2\, A{}     }{\left(A^2{} k^2+ B^2{} \right)} \frac1{(p-k)^4}\ ,\label{Sigma1}\\
\Sigma_2(p):=&
\frac{i}{\alpha }\int \dbar^4k\,
\frac{k^2\,  B{}  }{\left(A^2{} k^2+ B^2{} \right)} \frac1{ (p-k)^4}\ .
%-------------------------------------------------------------------------------------------------------------------------------------------------------% 
\label{Sigma2}
\end{align}
%-------------------------------------------------------------------------------------------------------------------------------------------------------%
$\Sigma_1(p)$ needs to be in a form in which $\gamma p$ appears explicitly such that
$A(p^2)$ can be easily read off (\ref{SD_eqn})   by comparing coefficients.
Multiply  $\Sigma_1$ by $\gamma p$ to obtain 
%-------------------------------------------------------------------------------------------------------------------------------------------------------% 
\begin{align}
\frac{1}{4} {\rm Tr}\,\gamma p \Sigma_1(p)
%-------------------------------------------------------------------------------------------------------------------------------------------------------% 
=&  \frac {-1}{2\alpha } \int \dbar^4k 
\frac{ k^2 A{} }{\left(A^2  k^2+ B^2  \right)}  
\nonumber\\&\times 
\bigg(
\frac{1 }{(p-k)^2 }
-
\frac{ p^2+k^2   }{ (p-k)^4}
\bigg)\ .
%-------------------------------------------------------------------------------------------------------------------------------------------------------% 
\label{SD_eqn_3}
\end{align}
%-------------------------------------------------------------------------------------------------------------------------------------------------------% 
Now multiply (\ref{SD_eqn_3}) by $\gamma p$, bearing in mind that 
by (\ref{wick_rot_3A})  $(\gamma p)^2=p^2$, to obtain
%-------------------------------------------------------------------------------------------------------------------------------------------------------% 
\begin{align}
p^2 \Sigma_1(p)=&\frac {-1}{2\alpha }\gamma p
\int \dbar^4k 
\frac{k^2  A }{\left(A^2 k^2+ B^2 \right)} 
\nonumber\\&\times \left(
\frac1{ (p-k)^2 }
-
\frac{  p^2+k^2     }{(p-k)^4}
\right)
%-------------------------------------------------------------------------------------------------------------------------------------------------------% 
\ .\label{SD_eqn_4}
\end{align}
% -------------------------------------------------------------------------------------------------------------------------------------------------------% 
In this form the $\gamma p$ term is explicit and  
(\ref{SD_eqn_2b}) reads
%-------------------------------------------------------------------------------------------------------------------------------------------------------% 
\begin{align}
i \Sigma(p)=&
\frac {-1}{2\alpha p^2} \gamma p 
\int \dbar^4k 
\frac{k^2  A  }{\left(A^2  k^2+ B^2  \right)} 
\left(
\frac1{ (p-k)^2 }
-
\frac{  p^2+k^2 }{ (p-k)^4}
\right)\nonumber\\
&+
\frac i{\alpha }\int \dbar^4k 
\frac{k^2   B   }{\left(A^2  k^2+ B^2  \right)} \frac1{ (p-k)^4}
\ .\label{SD_eqn_5}\end{align}\noindent
%-------------------------------------------------------------------------------------------------------------------------------------------------------% 
Substitute (\ref{SD_eqn_5}) back into (\ref{SD_eqn}) and compare coefficients on both sides to obtain the integral
equations
%-------------------------------------------------------------------------------------------------------------------------------------------------------% 
\begin{align}
A(p^2)=&1+
\frac 1{2\alpha p^2}  \int \dbar^4k 
\frac{k^2  A   }{\left(A^2  k^2+ B^2  \right)}
\nonumber\\&\times
\left(
\frac1{(p-k)^2}
-
\frac{  p^2+k^2   }{ (p-k)^4}
\right)
\ ,
\label{SD_eqn_6}
\\[1em]
%-------------------------------------------------------------------------------------------------------------------------------------------------------% 
B(p^2)=&
\frac 1{\alpha }\int \dbar^4k\,
\frac{k^2 B   }{\left(A^2 k^2+ B^2  \right)} \frac1{ (p-k)^4}
\ .\label{SD_eqn_7}\end{align}
\par 
%-------------------------------------------------------------------------------------------------------------------------------------------------------%
Write (\ref{SD_eqn_6}) and (\ref{SD_eqn_7})  in terms of
four-dimensional polar coordinates:
%-------------------------------------------------------------------------------------------------------------------------------------------------------% 
\begin{widetext}
{\small 
\begin{align}
A(p^2)=&1+
\frac {1}{2\alpha p^2}\frac1{(2\pi)^{4}}\int^\infty_0  dk\int^\pi_0  d\theta_1\int^{\pi}_0 d\theta_2 \int ^{2\pi}_0  d\theta_3 k^3\sin^2\theta_1\sin\theta_2
\frac{k^2 A  }{\left(A^2  k^2+ B^2  \right)}
\left(\frac1{ p^2+k^2-2p k\cos\theta_1  }
-\frac{p^2+k^2 }{(p^2+k^2-2p k\cos\theta_1 )^2}\right)
\nonumber\\[1em]
=&1+
\frac {\pi}{\alpha p^2}\frac1{(2\pi)^{4}}\int^\infty_0 dk^2\int^\pi_0 d\theta_1 \sin^2\theta_1
\frac{k^4 A   }{\left(A^2  k^2+ B^2  \right)}
\left(\frac1{ p^2+k^2-2p k\cos\theta_1  }
-\frac{p^2+k^2 }{(p^2+k^2-2p k\cos\theta_1 )^2}\right)
\ .\label{SD_eqn_8}\\[1em]
B(p^2)=&
\frac 1\alpha
\frac1{(2\pi)^{4}}\int^\infty_0 dk\int^\pi_0 d\theta_1\int^{\pi}_0d\theta_2\int ^{2\pi}_0 d\theta_3 k^3\sin^2\theta_1\sin\theta_2
\frac{k^2 B   }{\left(A^2  k^2+ B^2  \right)} \frac1{(p^2+k^2-2p k\cos\theta_1 )^2}
\nonumber\\[1em]
=&
\frac {2\pi}\alpha
\frac1{(2\pi)^{4}}\int^\infty_0 dk^2\int^\pi_0 d\theta_1\sin^2\theta_1
\frac{k^4 B  }{\left(A^2  k^2+ B^2  \right)} \frac1{(p^2+k^2-2p k\cos\theta_1 )^2}
\ .\label{SD_eqn_9}\end{align}
}\noindent
%-------------------------------------------------------------------------------------------------------------------------------------------------------% 
The $\theta_1$ integrals have the forms
%-------------------------------------------------------------------------------------------------------------------------------------------------------% 
\begin{align}
I_{1}=&
\int^\pi_0 d\theta_1\frac{\sin^2\theta_1}
{a-b\,\cos\theta_1 }
\ ,\label{angular_integral_1}\\
I_{ 2}=&
\int^\pi_0 d\theta_1\frac{\sin^2\theta_1}
{(a-b\,\cos\theta_1)^2}
\ ,\label{angular_integral_2}
\end{align}
%-------------------------------------------------------------------------------------------------------------------------------------------------------% 
where $a= p^2+k^2 >0$ and $b=2pk$,
such that 
(\ref{SD_eqn_8}) and (\ref{SD_eqn_9}) 
read
%-------------------------------------------------------------------------------------------------------------------------------------------------------% 
\begin{align}
A(p^2)
=&1+
\frac {\pi }{\alpha p^2 }\frac1{(2\pi)^{4}}\int^\infty_0 dk^2 
\frac{k^4  A   }{\left(A^2 k^2+ B^2  \right)}\,\left(I_{ 1}-(p^2+k^2 )I_{ 2}\right)
\ ,\label{SD_eqn_10A}\\[2em]
B(p^2)
=&
\frac {2\pi }{\alpha }\frac1{(2\pi)^4}\int^\infty_0 dk^2 
\frac{k^4  B  }{\left(A^2  k^2+ B^2  \right)}  I_{ 2}
\ .\label{SD_eqn_11A}\end{align}
%-------------------------------------------------------------------------------------------------------------------------------------------------------% 
The values of the integrals $I_{ 1}$ and $I_{ 2}$ are calculated in 
Appendix \S \ref{sec_angular_integral}.
Their final forms are given in Eqs.~(\ref{I1_eps_expansn})~and~(\ref{I2_eps_expansn}). 
%as expansions in powers of $p^2$.
Substitute them in (\ref{SD_eqn_10A})~and~(\ref{SD_eqn_11A})  to yield
%-------------------------------------------------------------------------------------------------------------------------------------------------------% 
\begin{align}
A(p^2)
=&1+\frac1{16 \pi ^2 \alpha p^2}
\int^\infty_0 dk^2 
\frac{A k^4}{ \left(A^2 k^2+B^2\right)}
\left[
\frac{p^2}{k^2(p^2-k^2)}\theta(k-p)+\frac{k^2}{p^2(k^2-p^2)}\theta(p-k)
\right]
\ \label{SD_eqn_11_B}\end{align}
%-------------------------------------------------------------------------------------------------------------------------------------------------------% 
and 
%-------------------------------------------------------------------------------------------------------------------------------------------------------% 
\begin{align}
B(p^2)
=&
\frac{1}{16\pi^2\alpha}\int^\infty_0 dk^2 
\frac{B k^4}{  \left(A^2 k^2+B^2\right)}
\left[
\frac1{k^2(k^2-p^2)}\theta(k-p)
+
\frac1{p^2(p^2-k^2)}\theta(p-k)
\right]
\ .\label{SD_eqn_11_C}\end{align}
%-------------------------------------------------------------------------------------------------------------------------------------------------------% 
Clearly the integrands in both expressions diverge at $k=p$.
However, 
the integration limits will be adjusted
to take values that agree with current experimental data, such that this
singular point will lie outside of the integration range.
This is 
elucidated in the next paragraph. 
The next step is to 
drop the 
contributions to 
the loop integrals in (\ref{SD_eqn_11_B}) and (\ref{SD_eqn_11_C})
from the region  $k<p$ while retaining 
just the pieces 
from the  region $k>p$.
This too is justified in the following paragraph.
To aid the discussion below
it is useful to expand the remaining terms in $A$ and $B$ in powers of $p^2$ 
to yield
%-------------------------------------------------------------------------------------------------------------------------------------------------------% 
\begin{align}
A(p^2)
=&1+\frac1{16 \pi ^2 \alpha }
\int^\infty_0 dk^2 
\frac{A }{ \left(A^2 k^2+B^2\right)}\left[
-1
-\frac{p^2 }{k^{2}}
-\frac{p^4}{k^{4}}\right]+O(p^6)
\ \label{SD_eqn_12}\end{align}
%-------------------------------------------------------------------------------------------------------------------------------------------------------% 
and 
%-------------------------------------------------------------------------------------------------------------------------------------------------------% 
\begin{align}
B(p^2)
=&
\frac{1}{16\pi^2\alpha}\int^\infty_0 dk^2 
\frac{B }{  \left(A^2 k^2+B^2\right)}
\left[1
+
\frac{p^2   }{k^2}
+
\frac{p^4 }{k^{4}}
\right]+O(p^6)
\ .\label{SD_eqn_13}\end{align}
%-------------------------------------------------------------------------------------------------------------------------------------------------------% 
\end{widetext}
%-------------------------------------------------------------------------------------------------------------------------------------------------------% 
\par
%-------------------------------------------------------------------------------------------------------------------------------------------------------% 
The integrals in (\ref{SD_eqn_11_B}) and (\ref{SD_eqn_11_C})
are 
ultraviolet divergent.
In contrast,  
notice that  the terms in the expansions in the integrals, (\ref{SD_eqn_12}) and (\ref{SD_eqn_12}),
except for the first term (independent of $p$) are infra-red divergent.
However,  in our scheme 
we implement an infra-red cutoff in addition to an ultraviolet cutoff. 
Physically the presence of this infra-red cutoff is needed in order to satisfy present experimental bounds on the existence of extra scalar excitations. 
From this bound we derive that the infra-red cutoff in the above integral should be of the order $\lambda \sim 1$ TeV.  
%-------------------------------------------------------------------------------------------------------------------------------------------------------% 
Evidently, the dominant contribution to $B(p^2)$ originates from the 
large $k$ region, provided that $\lambda\gg m_t$. The implication is that $B$ is log divergent.
The integral over $k^2$ is assumed to have an upper limit at $k=\Lambda$ with $\Lambda$ finite.
This essentially means that the  four momentum in the loop 
is bound from above (see Fig~\ref{fig_1}).
The upshot is that $\Lambda$ is  large to the extent that the value of the integral is saturated by the region where
$k$ is close to $\Lambda$.
In physical terms the loop is dominated by large values of $k$, or equivalently,
the loop is dominated by small distances of the order $1/\Lambda$,
with $\Lambda$ of the order of the Planck mass.
As a result it is reasonable to take $p$ to be of the order of $100$ GeV, close to the fermion (top-quark) mass.
Since 1 GeV is much smaller than the Planck scale, 
$p$ is effectively zero in the integral.\par 
%-------------------------------------------------------------------------------------------------------------------------------------------------------%
The leading order in $p^2$ contributions to $A$ and $B$ are straightforwardly read off
Eqs.~(\ref{SD_eqn_12}) and (\ref{SD_eqn_13}). By  
inserting a cut-off at the ultra-violet and at the infra-red end of the spectrum,
these LO contributions have the forms
%-------------------------------------------------------------------------------------------------------------------------------------------------------% 
\begin{align}
A_0
=&1-\frac1{16 \pi ^2 \alpha }
\int^{\Lambda^2}_{\lambda^2} dk^2 
\frac{A_0 }{ \left(A_0^2 k^2+B_0^2\right)}
\ ,\label{A_LO}\\
B_0
=&
\frac{1}{16\pi^2\alpha}\int^{\Lambda^2}_{\lambda^2} dk^2 
\frac{B_0 }{  \left(A_0^2 k^2+B_0^2\right)}
\ .\label{B_LO}\end{align}
%-------------------------------------------------------------------------------------------------------------------------------------------------------% 
By canceling 
a constant factor of $B_0$ on both sides of (\ref{B_LO})
and substituting a reparameterization of the form
$B_0=m_t A_0$, Eqs.~(\ref{A_LO}) and (\ref{B_LO})
reduce to
%-------------------------------------------------------------------------------------------------------------------------------------------------------% 
\begin{equation}
A_0
=1-\frac1{16 \pi ^2 \alpha }
\int^{\Lambda^2}_{\lambda^2} dk^2 
\frac{A_0 }{ A^2_0\left( k^2+m_t^2\right)}
\ ,\label{A_LO_1}\end{equation}
%-------------------------------------------------------------------------------------------------------------------------------------------------------% 
and 
%-------------------------------------------------------------------------------------------------------------------------------------------------------% 
\begin{equation}
1
=
\frac{1}{16\pi^2\alpha}\int^{\Lambda^2}_{\lambda^2} dk^2 
\frac{ 1}{ A_0^2 \left( k^2+m_t^2\right)}
\ ,\label{B_LO_1}\end{equation}
%-------------------------------------------------------------------------------------------------------------------------------------------------------% 
respectively.
Substitute (\ref{B_LO_1}) in (\ref{A_LO_1}) to obtain
%-------------------------------------------------------------------------------------------------------------------------------------------------------% 
\begin{equation}
A_0
=1-A_0
\qquad\Rightarrow\qquad A_0=\frac1{2}\ .\label{A_LO_2}\end{equation}
%-------------------------------------------------------------------------------------------------------------------------------------------------------% 
Now substitute (\ref{A_LO_2}) in (\ref{B_LO_1}) to find
%-------------------------------------------------------------------------------------------------------------------------------------------------------% 
\begin{equation}
1
=\frac{1}{4\pi^2\alpha}\ln\left( \frac{\Lambda^2 +m_t^2}{\lambda^2+m_t^2}\right)
\ .\label{B_LO_2}\end{equation}
%-------------------------------------------------------------------------------------------------------------------------------------------------------%
This gives rise to the existence of a critical value of the coupling constant 
%-------------------------------------------------------------------------------------------------------------------------------------------------------%
\begin{equation}\alpha_c = \frac{1}{4\pi^2}\ln\left( \frac{\Lambda^2 }{\lambda^2}\right) \sim 1.86641\ .
\label{alpha_c}
\end{equation}
%-------------------------------------------------------------------------------------------------------------------------------------------------------%
It can thus be inferred that for $\alpha < \alpha_c$ the fermion mass is generated dynamically. As  mentioned above, we assume that $\Lambda\gg \lambda \gg m_t$, 
where $\lambda$ is the infra-red cutoff of the momentum $k$ in the loop.
In this regime
%-------------------------------------------------------------------------------------------------------------------------------------------------------% 
\begin{equation}
m_t=\Lambda e^{-2\pi^2\alpha} \sqrt{1-e^{4\pi^2 (\alpha - \alpha_c)}}
\ .\label{B_LO_3}\end{equation}
%-------------------------------------------------------------------------------------------------------------------------------------------------------% 
It follows from (\ref{B_LO_3})  
that the generated top mass varies from $0$ at $\alpha = \alpha_c$ to a value that approaches $\Lambda$ 
at strong coupling $1/\alpha \gg 1$. All other intermediate values of $\alpha$ correspond to physical values of the fermion masses. 
For example, the top quark mass, which is approximately  $175$ GeV, is generated for $\alpha = 1.86564$. 
By comparing this value with the critical value of $\alpha$, we see that a certain type of the fine tuning is needed in the given model.  
\par
\begin{figure}[htb!]
% \centering
\includegraphics[width=9cm]{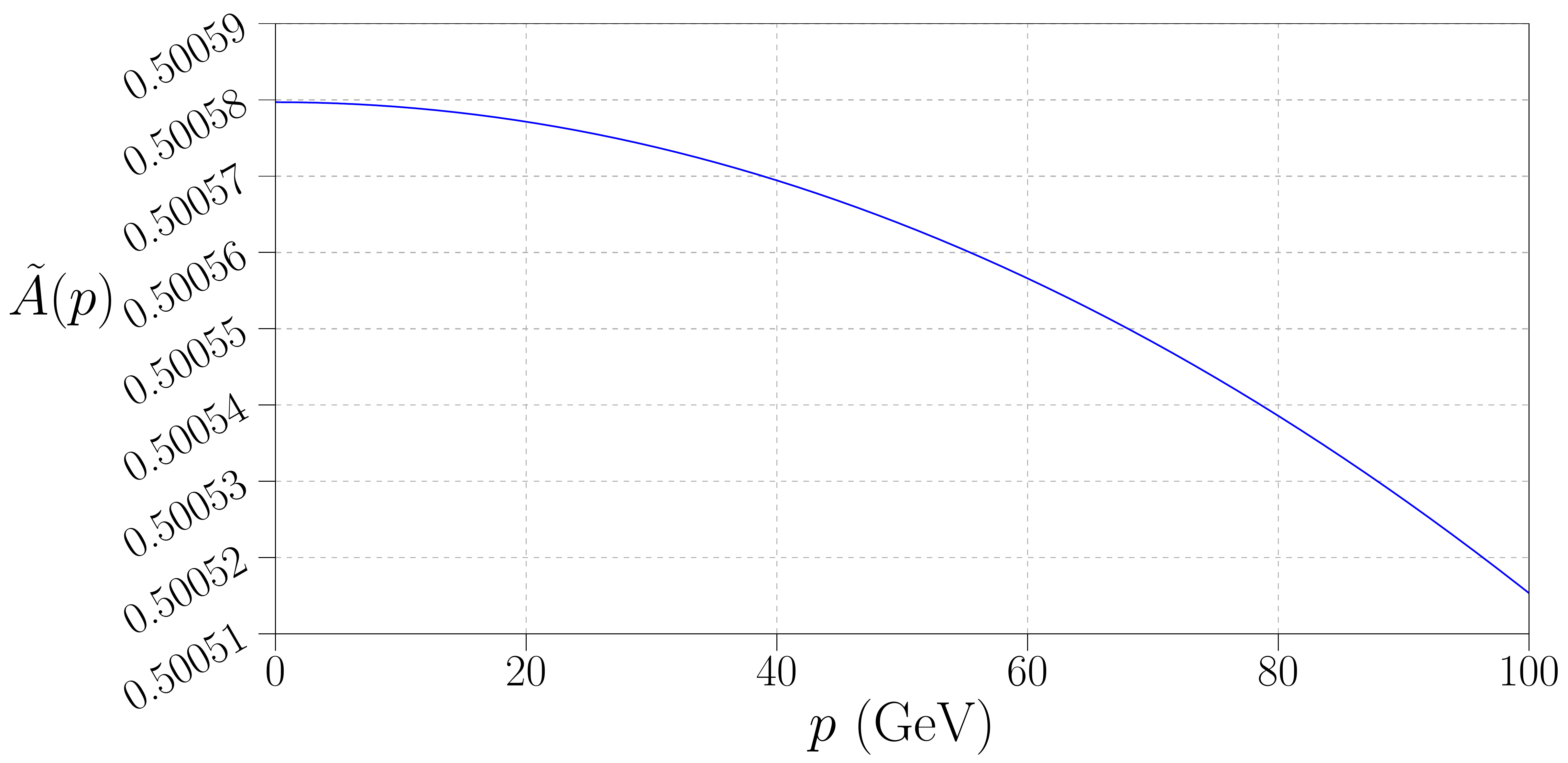}
\caption{\footnotesize 
The function $\tilde A(p)$ found by evaluating the integral in Eq.~(\ref{tilde_A}) 
for the values $\Lambda=10^{19}\,{\rm Gev}$ (the Planck mass),
$\lambda=1\,{\rm TeV}$,
$m_t=175\,{\rm GeV}$ and $\alpha=1.86564$.
The values of $\tilde A(p)$ are very close to $A_0=\frac1{2}$ as expected.
}
\label{fig_Aapprox}
\end{figure}

\begin{figure}[htb!]
% \centering
\includegraphics[width=9cm]{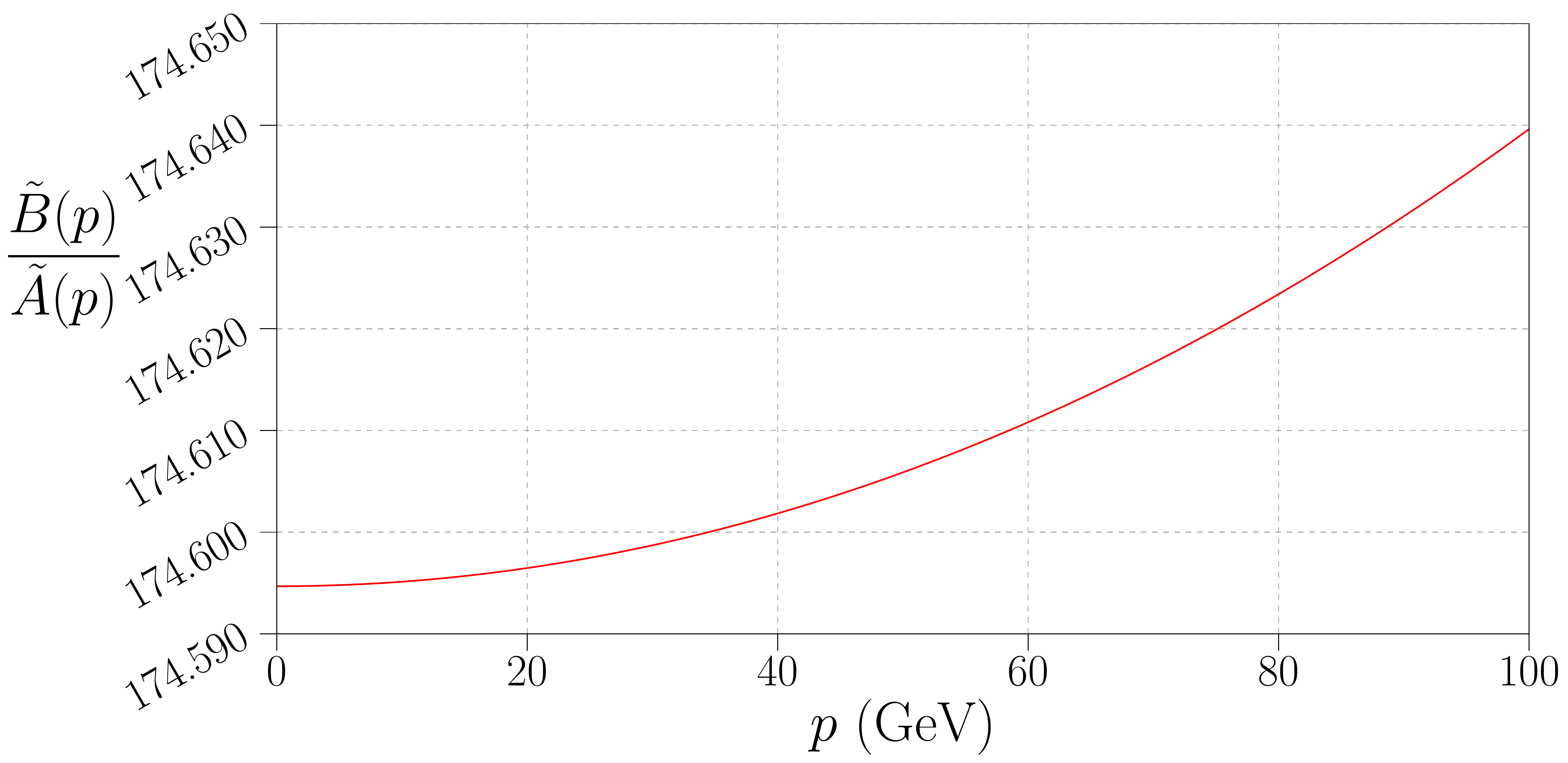}
\caption{\footnotesize 
The function $\tilde B(p)/\tilde A(p)$ found by evaluating the integral in
(\ref{tilde_B}) with the same numerical parameters as Fig.~\ref{fig_Aapprox}.
The values of $\tilde B(p)/\tilde A(p)$ are very close to $m_t=175$ GeV, as expected.
}
\label{fig_Bapprox}
\end{figure}

 
%-------------------------------------------------------------------------------------------------------------------------------------------------------% 
As a consistency check,
take the integrals in (\ref{SD_eqn_11_B}) and (\ref{SD_eqn_11_C})
but with the cutoffs $k^2=\lambda^2$ at the lower end  and $k^2=\Lambda^2$ at the upper end of the spectrum,
where recall that $\lambda\gg p $ is assumed  such that only the $k>p$ contribution survives.
Then 
substitute for $A$ and $B$ inside the integrands, the leading-order values found in (\ref{A_LO_2}) and (\ref{B_LO_3}).
Select the values $\Lambda=10^{19}\,{\rm Gev}$ (the Planck mass),
$\lambda=1\,{\rm TeV}$,
$m_t=175\,{\rm GeV}$ and $\alpha=1.86564$  then evaluate the integrals over $k^2$.
In this regime, the integrals
as functions of $p$ labeled as
$\tilde A(p)$ and $\tilde B(p)$,
have the forms
%-------------------------------------------------------------------------------------------------------------------------------------------------------%
\begin{align}
\tilde A(p)=&1+\frac1{16\pi^2\alpha}\int^{\Lambda^2}_{\lambda^2}dk^2\frac{A_0 k^4}{A_0^2 k^2+B_0^2}\frac{1}{(p^2-k^2)}\ ,
\label{tilde_A}\\
\tilde B(p)=& \frac1{16\pi^2\alpha}\int^{\Lambda^2}_{\lambda^2}dk^2\frac{B_0 k^2}{A_0^2 k^2+B_0^2}\frac{1}{(k^2-p^2)}\ ,
\label{tilde_B}
\end{align}
%-------------------------------------------------------------------------------------------------------------------------------------------------------%
where in Eqs.~(\ref{tilde_A}) and (\ref{tilde_B}),
$A_0=\frac1{2}$ and
$B_0=\frac{m_t}{2}=87.5$ GeV should be inserted.
The values
for $\tilde A(p)$ and $\tilde B(p)/\tilde A(p)$ as functions of $p$ are shown in the plots in Figs~\ref{fig_Aapprox} and Fig.~\ref{fig_Bapprox}.
The values of $\tilde A(p)$ are very close to $A_0=\frac1{2}$ as expected, while
the values of $\tilde B(p)/\tilde A(p)$ are very close to $m_t=175$ GeV, also as expected.
%-------------------------------------------------------------------------------------------------------------------------------------------------------%
%-------------------------------------------------------------------------------------------------------------------------------------------------------%
%-------------------------------------------------------------------------------------------------------------------------------------------------------%
% III.
% CONCLUSIONS
\section{Conclusion}
\label{sec_conclusion}
\noindent
%-------------------------------------------------------------------------------------------------------------------------------------------------------%
In this paper we have proposed a novel scheme that could pave the way 
for the construction of the ultra-violet completion of the SM.
The said mechanism, instead of 
the conventional Higgs boson field,
features a set of 36
zero-dimension scalar fields that give rise to
dynamical symmetry breaking.
In consequence 
we have presented a mechanism for 
obtaining fermion mass terms in the SM action.
We have 
implemented this mechanism to the top-quark sector of the action,
and shown that it is capable of predicting a value for the top quark mass
in agreement with the measured value, within
specific limits
imposed on the energy scale of the interactions between the scalars and the top-quark.
\par 
%-------------------------------------------------------------------------------------------------------------------------------------------------------%
These scalar fields are precisely the  couplings
between the fermions and the vielbein field of the background gravitational field.  
The form of the coupling we propose
is a direct extension of the non-minimal coupling of fermions to the vielbein
\cite{Alexandrov:2008iy},
in which  the constant coupling terms are promoted to fields in their own right.
We further generalize to a scenario where the coupling terms between the vielbein and 
the various different fermion flavours,
are distinctly
different fields for each different fermion in the SM action.
The outcome is that we obtain a total of  36 scalar fields 
that constitute the coupling terms.
At the same time
this  is the precise number of zero-dimension scalar fields
required to have a vanishing Weyl anomaly \cite{Boyle:2021jaz},
this being the case for the action we have formulated.
\par 
%-------------------------------------------------------------------------------------------------------------------------------------------------------%
But more fundamentally, in the model that we propose
the zero-dimension scalar fields have  manifestly geometrical origins. 
We assume that the sector of the theory containing these zero-dimension scalar fields,
is valid within the range of energy scales between $\lambda \sim 1 $ TeV and $\Lambda \sim m_P \sim 10^{19}$ GeV. 
The presence of the infrared cutoff, $\lambda$ is necessary in order to satisfy the currently known lower bound on extra scalar fields in the SM. 
The ultraviolet cutoff, $\Lambda$  is naturally of the order of the Plank mass,
the scale at which  quantum gravity effects  are expected to   play a significant role. 
\par 
%-------------------------------------------------------------------------------------------------------------------------------------------------------%
On grounds that the Weyl anomaly vanishes in the theory presented in this work,
then arguably this 
theory
offers a solution for the cosmological constant problem.
We argue that within this theory the fermion masses are 
generated dynamically due to the exchange of such scalars between the SM fermions. 
We have tested our model by implementing
a simplified toy-model version, containing just a doublet of the top quark and bottom quark 
interacting via the  exchange of the zero-dimension scalars. 
Specifically, we used an approach based on the truncated  Schwinger - Dyson equations (the so - called rainbow 
approximation), which
enables a description of dynamical symmetry breaking.%, at least  qualitatively. 
Our observations show that there exists a certain critical value of the coupling constant $\alpha$. For 
smaller values (strong coupling) 
the dynamical fermion mass is generated. In order for the fermion mass to be close to the top quark mass,  
the value of $\alpha$ must be close to the critical value.
\par 
%-------------------------------------------------------------------------------------------------------------------------------------------------------%
Thinking beyond the results in this article,
the toy model discussed here yields equal masses for both the top and bottom quarks. 
Without doubt this model 
should be generalized to incorporate different masses for the top and bottom quarks.
In particular
a prospective generalized model should have   a suppression of the bottom-quark mass in 
comparison with the top quark mass.
Above all, the interactions between other components of the scalars should be
taken into account, which give rise to the smaller masses of these fermions. 
Further afield the question arises of how to incorporate the generation of the difference in masses 
between the  $u$ and $d$ quarks, the  $c$ and $s$ quarks, 
as well as the difference in masses between the charged leptons and their neutrinos. 
This generalized theory is beyond the scope of this paper but will be addressed in a follow up paper.  
%-------------------------------------------------------------------------------------------------------------------------------------------------------%
\acknowledgements
This work has been supported by the European Research Council (ERC) under the European Union's Horizon 2020 research and innovation programme (Grant Agreement No. 694248).
\appendix
%-------------------------------------------------------------------------------------------------------------------------------------------------------%
\section{Wick rotations}\label{sec_wick_rotatns}\noindent 
%-------------------------------------------------------------------------------------------------------------------------------------------------------%
In Minkowski space the $\gamma$ matrices satisfy the Clifford algebra given by 
%-------------------------------------------------------------------------------------------------------------------------------------------------------%
\begin{equation}\{\gamma^\mu,\gamma^\nu\}=2\eta^{\mu\nu}\ ,\qquad (\mu,\nu=0,1,2,3)\ ,\label{wick_rot_1}\end{equation}
%-------------------------------------------------------------------------------------------------------------------------------------------------------%
where $\eta^{\mu\nu}={\rm diag}(1,-1,-1,-1)$.
A wick rotation comprises a transformation
of the $\gamma$ matrices %as defined in Minkowski space 
to new matrices, $\gamma_E$ defined by
%-------------------------------------------------------------------------------------------------------------------------------------------------------%
\begin{equation}\gamma^4_{E}=\gamma^0
\ ,\qquad \gamma^i_E =-i \gamma^i\ ,\qquad (i=1,2,3)\label{wick_rot_2A}\end{equation}
%-------------------------------------------------------------------------------------------------------------------------------------------------------%
such that 
%-------------------------------------------------------------------------------------------------------------------------------------------------------%
\begin{equation}\{\gamma_{E}^\mu,\gamma_{E}^\nu\}=2\delta^{\mu\nu}\ ,\qquad (\mu,\nu=1,2,3,4).\label{wick_rot_3A}\end{equation}
%-------------------------------------------------------------------------------------------------------------------------------------------------------%
Four-vector components get transformed under a Wick rotation to new components. For example given components $k^\mu$
of the four vector $k$ in Minkowski space,
%-------------------------------------------------------------------------------------------------------------------------------------------------------%
\begin{equation}k^4_{E}=ik^0
\ ,\qquad k^i_{E} =k^i\ ,\qquad (i=1,2,3)\ .\label{wick_rot_4}\end{equation}
%-------------------------------------------------------------------------------------------------------------------------------------------------------%
The implication is that 
%-------------------------------------------------------------------------------------------------------------------------------------------------------%
\begin{widetext}\begin{equation}k^2=\left(k^0\right)^2-\left(k^i\right)^2=
-\left(k_{E}^4\right)^2
-\left(k_{E}^1\right)^2-\left(k_{E}^2\right)^2-\left(k_{E}^3\right)^2
\equiv-k_{E}^2\ .
\label{wick_rot_5}\end{equation}
%-------------------------------------------------------------------------------------------------------------------------------------------------------%
\begin{equation}
\gamma k=
\gamma^0 k^0-\gamma^i k_i
=
-i\gamma^4_{{E}}k^4_{{E}}
-i\gamma^1_{{E}}k^1_{{E}}-i\gamma^2_{{E}}k^2_{{E}}-i\gamma^3_{{E}}k^3_{{E}}\equiv -i\gamma_{{E}}k_{{E}}\ .
\label{wick_rot_6A}\end{equation}
\end{widetext}
%-------------------------------------------------------------------------------------------------------------------------------------------------------%
\section{Angular integral}\label{sec_angular_integral}\noindent
%-------------------------------------------------------------------------------------------------------------------------------------------------------%
The integrals  in (\ref{angular_integral_1}) and (\ref{angular_integral_2}) are solved in this appendix,
whose forms are
%-------------------------------------------------------------------------------------------------------------------------------------------------------%
\begin{equation}\begin{aligned}
I_{ 1} =&\int^\pi_0d\theta\,\frac{\sin^2\theta}{\left(a-b\cos\theta\right) }\ ,\\  
I_{ 2} =&\int^\pi_0d\theta\,\frac{\sin^2\theta}{\left(a-b\cos\theta\right)^2}\ ,
\end{aligned}\label{angular_integral_1}\end{equation}
%-------------------------------------------------------------------------------------------------------------------------------------------------------%
where
%-------------------------------------------------------------------------------------------------------------------------------------------------------%
\begin{equation}
a=p^2+k^2\ ,\qquad b=2  p  k\ .\label{angular_integral_1A}
\end{equation}
%-------------------------------------------------------------------------------------------------------------------------------------------------------%
They both have the same structure, namely
%-------------------------------------------------------------------------------------------------------------------------------------------------------%
\begin{equation}
I_{ n} = \int^\pi_0d\theta\,\frac{\sin^2\theta}{\left(a-b\cos\theta\right)^n}\ ,\label{angular_integral_1B}
\end{equation}
%-------------------------------------------------------------------------------------------------------------------------------------------------------%
and are evaluated in the same way.\par
%-------------------------------------------------------------------------------------------------------------------------------------------------------%
First write it as an integral from $0$ to $2\pi$ so that it can be converted into a contour integral along a
closed circlular path in the complex plane.
Since $\cos\theta=\cos(2\pi-\theta)$ and $\sin^2\theta=\sin^2(2\pi-\theta)$
then $I_{ n} $ can equally be written as
%-------------------------------------------------------------------------------------------------------------------------------------------------------%
\begin{align}
I_{ n} 
=&\int^{\pi}_{0}d\theta\,\frac{\sin^2(2\pi-\theta)}{\left(a-b\cos(2\pi-\theta)\right)^n}
%\nonumber\\
% =&-\int^{\pi}_{2\pi}d\xi\,\frac{\sin^2\xi}{\left(a-b\cos\xi\right)^n}
\nonumber\\
=&\int^{2\pi}_{\pi}d\xi\,\frac{\sin^2\xi}{\left(a-b\cos\xi\right)^n}\ ,\label{angular_integral_2}
\end{align}
%-------------------------------------------------------------------------------------------------------------------------------------------------------%
such that
%-------------------------------------------------------------------------------------------------------------------------------------------------------%
\begin{equation}
I_{ n} =\frac1 {2}\int^{2\pi}_0d\theta\,\frac{\sin^2\theta}{\left(a-b\cos\theta\right)^n}\ .\label{angular_integral_3}
\end{equation}
%-------------------------------------------------------------------------------------------------------------------------------------------------------%
Now write this as a closed integral over the variable $z=e^{i\theta}$ in the complex plane,
with $\sin\theta=(-i/2)(z-z^{-1})$ and
$\cos\theta=(1/2)(z+z^{-1}) $:
%-------------------------------------------------------------------------------------------------------------------------------------------------------%
\begin{align}
I_{ n}
=&
\frac i{8}\oint dz\,\frac{2^nz^{n-3}\left(z^2-1\right)^2}
{\left(2a z-b z^2-b \right)^n}
\ .\label{angular_integral_4}
\end{align}
%-------------------------------------------------------------------------------------------------------------------------------------------------------%
Write the denominator of the integrand as
%-------------------------------------------------------------------------------------------------------------------------------------------------------%
\begin{equation}
\left(2a z-b z^2-b \right) ^n
=
(-b)^n\left( z-z_1\right)^n\left( z-z_2\right)^n\ ,
\label{angular_integral_5}
\end{equation}
%-------------------------------------------------------------------------------------------------------------------------------------------------------%
where 
%-------------------------------------------------------------------------------------------------------------------------------------------------------%
\begin{equation}
z_1 =\frac{a+\sqrt{ a^2-b^2}}{b}\ ,
\qquad
z_2=\frac{a-\sqrt{ a^2-b^2}}{b}\ ,
\label{angular_integral_6}
\end{equation}
%-------------------------------------------------------------------------------------------------------------------------------------------------------%
to bring the integral into the form
%-------------------------------------------------------------------------------------------------------------------------------------------------------%
\begin{equation}
I_{ n} =2^{n-3}i \oint dz\,\frac{z^{n-3}\left(z^2-1\right)^2}
{(-b)^n\left( z-z_1\right)^n\left( z-z_2\right)^n}
\ .\label{angular_integral_7}
\end{equation}
%-------------------------------------------------------------------------------------------------------------------------------------------------------%
Now the integral has the familiar form $\displaystyle \oint dz\,\dfrac{f(z)}{(z-z_1)^n(z-z_2)^m}$
where the contour is the unit circle
with $f(z)$ analytic and continuous at $z_1$ and $z_2$,
and it can be solved  by the standard method of summing over the residues of $f(z)$.
\par 
%-------------------------------------------------------------------------------------------------------------------------------------------------------%
% First the $n=4$ is solved.
Note carefully the location of the poles. From (\ref{angular_integral_1A})
it follows that 
%-------------------------------------------------------------------------------------------------------------------------------------------------------%
\begin{equation}a-b=p^2+k^2 -2pk=(p-k)^2 >0\ , \label{angular_integral_7A}\end{equation}
%-------------------------------------------------------------------------------------------------------------------------------------------------------%
and there it is always true that
%-------------------------------------------------------------------------------------------------------------------------------------------------------%
\begin{equation}a>b  \ .\label{angular_integral_7B} \end{equation}
%-------------------------------------------------------------------------------------------------------------------------------------------------------%
Accordingly 
%-------------------------------------------------------------------------------------------------------------------------------------------------------%
\begin{align}
z_1 =&
\frac k{p} \theta(k-p)
+\frac p{k}
\theta(p-k)
\ ,
\\
z_2 =& 
\frac p{k} \theta(k-p)
+\frac k{p}
\theta(p-k)
\ .
\end{align} 
%-------------------------------------------------------------------------------------------------------------------------------------------------------%
In conclusion
$|z_2|<1$ while $|z_1|>1$,
so the pole at $z_1$ is discounted because it is not inside the unit circle.
This leaves two poles in the expression 
(\ref{angular_integral_7}): one at $z=0$ and one at $z=z_2$.
\par 
%-------------------------------------------------------------------------------------------------------------------------------------------------------%
\begin{widetext}
For $n=2$ the integral in  (\ref{angular_integral_7}), by the Cauchy formula, has the form
%-------------------------------------------------------------------------------------------------------------------------------------------------------%
\begin{align}
I_{ 2} =&
\frac i{2}(2\pi i)\frac1{b^2}\left[
\left( 
\frac{\left(z^2-1\right)^2}{ \left( z-z_1\right)^2\left( z-z_2\right)^2}\right)\bigg|_{z=0}
+
\frac d{dz}\left( 
\frac{\left(z^2-1\right)^2}{ z\left( z-z_1\right)^2  }\right)\bigg|_{z=z_2}
\right]
\nonumber\\
=&
- \frac\pi{(2pk)^2}\left[ 
\frac1{ \left( z_1 z_2\right)^2}
-\frac{\left(z_2^2-1\right)^2}{z_2^2 (z_2-z_1)^2}-\frac{2 \left(z_2^2-1\right)^2}{z_2 (z_2-z_1)^3}+\frac{4 \left(z_2^2-1\right)}{(z_2-z_1)^2}
\right]
\ ,\label{angular_integral_10}
\end{align}
%-------------------------------------------------------------------------------------------------------------------------------------------------------%
while for $n=1$ it is
%-------------------------------------------------------------------------------------------------------------------------------------------------------%
\begin{align}
I_{ 1} =&
\frac i{4}(2\pi i)\frac1{(-b)}\left[
\frac d{dz}
\left( 
\frac{\left(z^2-1\right)^2}{ \left( z-z_1\right) \left( z-z_2\right) }\right)\bigg|_{z=0}
+
\left( 
\frac{\left(z^2-1\right)^2}{ z^2\left( z-z_1\right) }\right)\bigg|_{z=z_2}
\right]
\nonumber\\
=&
\frac\pi{4pk }\left[ 
\frac{{z_1}+{z_2}}{{z_1}^2 {z_2}^2}
+
\frac{\left({z_2}^2-1\right)^2}{{z_2}^2 ({z_2}-{z_1})}
\right]
\ .\label{angular_integral_10}
\end{align}
%-------------------------------------------------------------------------------------------------------------------------------------------------------%
Finally
substitute the explicit forms of $z_1$ and $z_2$ to obtain
%-------------------------------------------------------------------------------------------------------------------------------------------------------%
\begin{align}
I_1=& 
\frac{\pi }{2 k^2}
\theta(k-p)
+
\frac{\pi }{2 p^2}
\theta(p-k) \ ,
\label{I1_eps_expansn}\end{align}
%-------------------------------------------------------------------------------------------------------------------------------------------------------%
and 
%-------------------------------------------------------------------------------------------------------------------------------------------------------%
\begin{align}
I_2=&-\frac{\pi  }{2k^2(p^2-k^2)} \theta(k-p)
-\frac\pi{ 2p^2(k^2-p^2)}
\theta(p-k)\ .
\label{I2_eps_expansn}\end{align}
%-------------------------------------------------------------------------------------------------------------------------------------------------------%
\end{widetext}
%-------------------------------------------------------------------------------------------------------------------------------------------------------%
\bibliographystyle{unsrt}
\bibliography{references}

\begin{thebibliography}{10}

\bibitem{Boyle:2021jaz}
Latham Boyle and Neil Turok.
\newblock {\it ``Cancelling the vacuum energy and Weyl anomaly in the standard
  model with dimension-zero scalar fields''}.
\newblock \url{https://arxiv.org/pdf/2110.06258.pdf}, 2021.

\bibitem{Alexandrov:2008iy}
Sergei Alexandrov.
\newblock {Immirzi parameter and fermions with non-minimal coupling}.
\newblock {\em Class. Quant. Grav.}, 25:145012, 2008.

\bibitem{Higgs:1964ia}
Peter~W. Higgs.
\newblock {Broken symmetries, massless particles and gauge fields}.
\newblock {\em Phys. Lett.}, 12:132--133, 1964.

\bibitem{Higgs:1966ev}
Peter~W. Higgs.
\newblock {Spontaneous Symmetry Breakdown without Massless Bosons}.
\newblock {\em Phys. Rev.}, 145:1156--1163, 1966.

\bibitem{Kibble:1967sv}
T.~W.~B. Kibble.
\newblock {Symmetry breaking in nonAbelian gauge theories}.
\newblock {\em Phys. Rev.}, 155:1554--1561, 1967.

\bibitem{Englert:1964et}
F.~Englert and R.~Brout.
\newblock {Broken Symmetry and the Mass of Gauge Vector Mesons}.
\newblock {\em Phys. Rev. Lett.}, 13:321--323, 1964.

\bibitem{Guralnik:1964eu}
G.~S. Guralnik, C.~R. Hagen, and T.~W.~B. Kibble.
\newblock {Global Conservation Laws and Massless Particles}.
\newblock {\em Phys. Rev. Lett.}, 13:585--587, 1964.

\bibitem{Weinberg:1967kj}
Steven Weinberg.
\newblock {Precise relations between the spectra of vector and axial vector
  mesons}.
\newblock {\em Phys. Rev. Lett.}, 18:507--509, 1967.

\bibitem{Glashow:1961tr}
S.~L. Glashow.
\newblock {Partial Symmetries of Weak Interactions}.
\newblock {\em Nucl. Phys.}, 22:579--588, 1961.

\bibitem{Weinberg:1967tq}
Steven Weinberg.
\newblock {A Model of Leptons}.
\newblock {\em Phys. Rev. Lett.}, 19:1264--1266, 1967.

\bibitem{salam68}
A.~Salam.
\newblock {\it\,``Relativistic Groups and Analyticity''\,}.
\newblock In {\em {8th Nobel Symposium on Elementary Particle Theory\, ,
  1968}}, page 367, 1968.

\bibitem{Politzer:1973fx}
H.~David Politzer.
\newblock {Reliable Perturbative Results for Strong Interactions?}
\newblock {\em Phys. Rev. Lett.}, 30:1346--1349, 1973.

\bibitem{Cahn:1989by}
R.~N. Cahn and G.~Goldhaber.
\newblock {\em {The experimental foundations of particle physics}}.
\newblock Cambridge Univ. Press, Cambridge, 2009.

\bibitem{Fritzsch:1973pi}
H.~Fritzsch, Murray Gell-Mann, and H.~Leutwyler.
\newblock {Advantages of the Color Octet Gluon Picture}.
\newblock {\em Phys. Lett. B}, 47:365--368, 1973.

\bibitem{Nakada:1958zz}
M.~P. Nakada, J.~D. Anderson, C.~C. Gardner, J.~McClure', and C.~Wong.
\newblock {Neutron Spectrum from p+d Reaction}.
\newblock {\em Phys. Rev.}, 110:594--595, 1958.

\bibitem{Nambu:1960tm}
Yoichiro Nambu.
\newblock {Quasiparticles and Gauge Invariance in the Theory of
  Superconductivity}.
\newblock {\em Phys. Rev.}, 117:648--663, 1960.

\bibitem{Schwinger:1962tn}
Julian~S. Schwinger.
\newblock {Gauge Invariance and Mass}.
\newblock {\em Phys. Rev.}, 125:397--398, 1962.

\bibitem{CMS:2012bfw}
Serguei Chatrchyan et~al.
\newblock {Search for the Standard Model Higgs Boson Produced in Association
  with $W$ and $Z$ Bosons in $pp$ Collisions at $\sqrt{s}=7$ TeV}.
\newblock {\em JHEP}, 11:088, 2012.

\bibitem{Mariotti:2015psa}
Chiara Mariotti.
\newblock {Observation of a new Boson at a Mass of 125 gev With the CMS
  Experiment at the LHC}.
\newblock In {\em {13th Marcel Grossmann Meeting on Recent Developments in
  Theoretical and Experimental General Relativity, Astrophysics, and
  Relativistic Field Theories}}, pages 352--372, 2015.

\bibitem{CMS:2012qbp}
Serguei Chatrchyan et~al.
\newblock {Observation of a New Boson at a Mass of 125 GeV with the CMS
  Experiment at the LHC}.
\newblock {\em Phys. Lett. B}, 716:30--61, 2012.

\bibitem{CMS:2012dun}
Serguei Chatrchyan et~al.
\newblock {A Search for a Doubly-Charged Higgs Boson in $pp$ Collisions at
  $\sqrt{s}=7$ TeV}.
\newblock {\em Eur. Phys. J. C}, 72:2189, 2012.

\bibitem{ATLAS:2012yve}
Georges Aad et~al.
\newblock {Observation of a new particle in the search for the Standard Model
  Higgs boson with the ATLAS detector at the LHC}.
\newblock {\em Phys. Lett. B}, 716:1--29, 2012.

\bibitem{Mountricha:2012cja}
Eleni Mountricha.
\newblock {\em {Search for the Standard Model Higgs boson in the H
  \textrightarrow{} ZZ(\ensuremath{*}) \textrightarrow{} 4\ensuremath{\ell}
  channel with the ATLAS experiment at the LHC leading to the observation of a
  new particle compatible with the Higgs boson}}.
\newblock PhD thesis, Natl. Tech. U., Athens and Orsay, 10 2012.

\bibitem{Lane:2002wv}
Kenneth Lane.
\newblock {\it ``Two Lectures on Technicolor''}.
\newblock \url{https://arxiv.org/abs/hep-ph/0202255}, 2002.

\bibitem{Terazawa:1980nck}
H.~Terazawa.
\newblock {t quark mass predicted from a sum rule for lepton and quark masses}.
\newblock {\em Phys. Rev. D}, 22:2921--2921, 1980.
\newblock [Erratum: Phys.Rev.D 41, 3541 (1990)].

\bibitem{Terazawa:1976eq}
H.~Terazawa, Y.~Chikashige, K.~Akama, and T.~Matsuki.
\newblock {Simple Relation Between the Fine Structure and Gravitational
  Constants}.
\newblock {\em Phys. Rev. D}, 15:1181, 1977.

\bibitem{Bardeen:1989ds}
William~A. Bardeen, Christopher~T. Hill, and Manfred Lindner.
\newblock {Minimal Dynamical Symmetry Breaking of the Standard Model}.
\newblock {\em Phys. Rev. D}, 41:1647, 1990.

\bibitem{Marciano:1989xd}
W.~J. Marciano.
\newblock {HEAVY TOP QUARK MASS PREDICTIONS}.
\newblock {\em Phys. Rev. Lett.}, 62:2793--2796, 1989.

\bibitem{Hill:1991at}
Christopher~T. Hill.
\newblock {Topcolor: Top quark condensation in a gauge extension of the
  standard model}.
\newblock {\em Phys. Lett. B}, 266:419--424, 1991.

\bibitem{Hill:2002ap}
Christopher~T. Hill and Elizabeth~H. Simmons.
\newblock {Strong Dynamics and Electroweak Symmetry Breaking}.
\newblock {\em Phys. Rept.}, 381:235--402, 2003.
\newblock [Erratum: Phys.Rept. 390, 553--554 (2004)].

\bibitem{Miransky:1989ds}
V.~A. Miransky, Masaharu Tanabashi, and Koichi Yamawaki.
\newblock {Is the t Quark Responsible for the Mass of W and Z Bosons?}
\newblock {\em Mod. Phys. Lett. A}, 4:1043, 1989.

\bibitem{Chivukula:1998wd}
R.~Sekhar Chivukula, Bogdan~A. Dobrescu, Howard Georgi, and Christopher~T.
  Hill.
\newblock {Top Quark Seesaw Theory of Electroweak Symmetry Breaking}.
\newblock {\em Phys. Rev. D}, 59:075003, 1999.

\bibitem{Dobrescu:1997nm}
Bogdan~A. Dobrescu and Christopher~T. Hill.
\newblock {Electroweak symmetry breaking via top condensation seesaw}.
\newblock {\em Phys. Rev. Lett.}, 81:2634--2637, 1998.

\bibitem{Cheng:2013qwa}
Hsin-Chia Cheng, Bogdan~A. Dobrescu, and Jiayin Gu.
\newblock {Higgs Mass from Compositeness at a Multi-TeV Scale}.
\newblock {\em JHEP}, 08:095, 2014.

\bibitem{Susskind:1978ms}
Leonard Susskind.
\newblock {Dynamics of Spontaneous Symmetry Breaking in the Weinberg-Salam
  Theory}.
\newblock {\em Phys. Rev. D}, 20:2619--2625, 1979.

\bibitem{Weinberg:1975gm}
Steven Weinberg.
\newblock {Implications of Dynamical Symmetry Breaking}.
\newblock {\em Phys. Rev. D}, 13:974--996, 1976.
\newblock [Addendum: Phys.Rev.D 19, 1277--1280 (1979)].

\bibitem{Dimopoulos:1979es}
Savas Dimopoulos and Leonard Susskind.
\newblock {Mass Without Scalars}.
\newblock {\em Nucl. Phys. B}, 155:237--252, 1979.

\bibitem{Eichten:1979ah}
Estia Eichten and Kenneth~D. Lane.
\newblock {Dynamical Breaking of Weak Interaction Symmetries}.
\newblock {\em Phys. Lett. B}, 90:125--130, 1980.

\bibitem{Appelquist:1998rb}
Thomas Appelquist, Anuradha Ratnaweera, John Terning, and L.~C.~R.
  Wijewardhana.
\newblock {The Phase structure of an SU(N) gauge theory with N(f) flavors}.
\newblock {\em Phys. Rev. D}, 58:105017, 1998.

\bibitem{Gudnason:2006ug}
Sven~Bjarke Gudnason, Chris Kouvaris, and Francesco Sannino.
\newblock {Towards working technicolor: Effective theories and dark matter}.
\newblock {\em Phys. Rev. D}, 73:115003, 2006.

\bibitem{Lane:1993wz}
Kenneth~D. Lane.
\newblock {An Introduction to technicolor}.
\newblock In {\em {Theoretical Advanced Study Institute (TASI 93) in Elementary
  Particle Physics: The Building Blocks of Creation - From Microfermius to
  Megaparsecs}}, 6 1993.

\bibitem{Dugan:1984hq}
Michael~J. Dugan, Howard Georgi, and David~B. Kaplan.
\newblock {Anatomy of a Composite Higgs Model}.
\newblock {\em Nucl. Phys. B}, 254:299--326, 1985.

\bibitem{Perelstein:2005ka}
Maxim Perelstein.
\newblock {Little Higgs models and their phenomenology}.
\newblock {\em Prog. Part. Nucl. Phys.}, 58:247--291, 2007.

\bibitem{Arkani-Hamed:2002ikv}
N.~Arkani-Hamed, A.~G. Cohen, E.~Katz, and A.~E. Nelson.
\newblock {The Littlest Higgs}.
\newblock {\em JHEP}, 07:034, 2002.

\bibitem{Low:2002ws}
Ian Low, Witold Skiba, and David Tucker-Smith.
\newblock {Little Higgses from an antisymmetric condensate}.
\newblock {\em Phys. Rev. D}, 66:072001, 2002.

\bibitem{Berger:2005ht}
C.~F. Berger, M.~Perelstein, and F.~Petriello.
\newblock {Top quark properties in little Higgs models}.
\newblock In {\em {2005 International Linear Collider Physics and Detector
  Workshop and 2nd ILC Accelerator Workshop}}, 12 2005.

\bibitem{Redi:2012ha}
Michele Redi and Andrea Tesi.
\newblock {Implications of a Light Higgs in Composite Models}.
\newblock {\em JHEP}, 10:166, 2012.

\bibitem{Arhrib:2020tqk}
Abdesslam Arhrib, Rachid Benbrik, Hicham Harouiz, Stefano Moretti, Yan Wang,
  and Qi-Shu Yan.
\newblock {Implications of a light charged Higgs boson at the LHC run III in
  the 2HDM}.
\newblock {\em Phys. Rev. D}, 102(11):115040, 2020.

\bibitem{Kaplan:1991dc}
David~B. Kaplan.
\newblock {Flavor at SSC energies: A New mechanism for dynamically generated
  fermion masses}.
\newblock {\em Nucl. Phys. B}, 365:259--278, 1991.

\bibitem{Redi:2011zi}
Michele Redi and Andreas Weiler.
\newblock {Flavor and CP Invariant Composite Higgs Models}.
\newblock {\em JHEP}, 11:108, 2011.

\bibitem{Hong:2016uou}
Deog~Ki Hong and Du~Hwan Kim.
\newblock {Composite (pseudo) scalar contributions to muon g \ensuremath{-} 2}.
\newblock {\em Phys. Lett. B}, 758:370--372, 2016.

\bibitem{Matsuzaki:2016joz}
Shinya Matsuzaki and Koichi Yamawaki.
\newblock {Walking from 750 GeV to 950 GeV in the technipion zoo}.
\newblock {\em Phys. Rev. D}, 93(11):115027, 2016.

\bibitem{Harigaya:2016pnu}
Keisuke Harigaya and Yasunori Nomura.
\newblock {A Composite Model for the 750 GeV Diphoton Excess}.
\newblock {\em JHEP}, 03:091, 2016.

\bibitem{Nakai:2015ptz}
Yuichiro Nakai, Ryosuke Sato, and Kohsaku Tobioka.
\newblock {Footprints of New Strong Dynamics via Anomaly and the 750 GeV
  Diphoton}.
\newblock {\em Phys. Rev. Lett.}, 116(15):151802, 2016.

\bibitem{Franceschini:2015kwy}
Roberto Franceschini, Gian~F. Giudice, Jernej~F. Kamenik, Matthew McCullough,
  Alex Pomarol, Riccardo Rattazzi, Michele Redi, Francesco Riva, Alessandro
  Strumia, and Riccardo Torre.
\newblock {What is the $\gamma \gamma$ resonance at 750 GeV?}
\newblock {\em JHEP}, 03:144, 2016.

\bibitem{Molinaro:2015cwg}
Emiliano Molinaro, Francesco Sannino, and Natascia Vignaroli.
\newblock {Minimal Composite Dynamics versus Axion Origin of the Diphoton
  excess}.
\newblock {\em Mod. Phys. Lett. A}, 31(26):1650155, 2016.

\bibitem{Bian:2015kjt}
Ligong Bian, Ning Chen, Da~Liu, and Jing Shu.
\newblock {Hidden confining world on the 750 GeV diphoton excess}.
\newblock {\em Phys. Rev. D}, 93(9):095011, 2016.

\bibitem{Bai:2015nbs}
Yang Bai, Joshua Berger, and Ran Lu.
\newblock {750 GeV dark pion: Cousin of a dark G -parity odd WIMP}.
\newblock {\em Phys. Rev. D}, 93(7):076009, 2016.

\bibitem{Cline:2015msi}
James~M. Cline and Zuowei Liu.
\newblock {LHC diphotons from electroweakly pair-produced composite
  pseudoscalars}.
\newblock 12 2015.

\bibitem{Ko:2016wce}
P.~Ko and Takaaki Nomura.
\newblock {Dark sector shining through 750 GeV dark Higgs boson at the LHC}.
\newblock {\em Phys. Lett. B}, 758:205--211, 2016.

\bibitem{Redi:2016kip}
Michele Redi, Alessandro Strumia, Andrea Tesi, and Elena Vigiani.
\newblock {Di-photon resonance and Dark Matter as heavy pions}.
\newblock {\em JHEP}, 05:078, 2016.

\bibitem{Harigaya:2016eol}
Keisuke Harigaya and Yasunori Nomura.
\newblock {Hidden Pion Varieties in Composite Models for Diphoton Resonances}.
\newblock {\em Phys. Rev. D}, 94(7):075004, 2016.

\bibitem{Foot:2016llc}
Robert Foot and John Gargalionis.
\newblock {Explaining the 750 GeV diphoton excess with a colored scalar charged
  under a new confining gauge interaction}.
\newblock {\em Phys. Rev. D}, 94(1):011703, 2016.

\bibitem{Iwamoto:2016ral}
Sho Iwamoto, Gabriel Lee, Yael Shadmi, and Robert Ziegler.
\newblock {Diphoton Signals from Colorless Hidden Quarkonia}.
\newblock {\em Phys. Rev. D}, 94(1):015003, 2016.

\bibitem{Bai:2016vca}
Yang Bai, Joshua Berger, James Osborne, and Ben~A. Stefanek.
\newblock {Phenomenology of Strongly Coupled Chiral Gauge Theories}.
\newblock {\em JHEP}, 11:153, 2016.

\bibitem{Kobakhidze:2015ldh}
Archil Kobakhidze, Fei Wang, Lei Wu, Jin~Min Yang, and Mengchao Zhang.
\newblock {750 GeV diphoton resonance in a top and bottom seesaw model}.
\newblock {\em Phys. Lett. B}, 757:92--96, 2016.

\bibitem{Holdom:1988gs}
Bob Holdom.
\newblock {Raising Condensates Beyond the Ladder}.
\newblock {\em Phys. Lett. B}, 213:365--369, 1988.

\bibitem{Holdom:1989aa}
B.~Holdom.
\newblock {HOLDOM REPLIES TO: COMMENT ON `CONTINUUM LIMIT OF QUENCHED
  THEORIES.'}.
\newblock {\em Phys. Rev. Lett.}, 63:1889, 1989.

\bibitem{Yamawaki:1996vr}
Koichi Yamawaki.
\newblock {Dynamical symmetry breaking with large anomalous dimension}.
\newblock In {\em {14th Symposium on Theoretical Physics: Dynamical Symmetry
  Breaking and Effective Field Theory}}, 3 1996.

\bibitem{Fukano:2010yv}
Hidenori~S. Fukano and Francesco Sannino.
\newblock {Conformal Window of Gauge Theories with Four-Fermion Interactions
  and Ideal Walking}.
\newblock {\em Phys. Rev. D}, 82:035021, 2010.

\bibitem{Rantaharju:2017eej}
Jarno Rantaharju, Claudio Pica, and Francesco Sannino.
\newblock {Ideal Walking Dynamics via a Gauged NJL Model}.
\newblock {\em Phys. Rev. D}, 96(1):014512, 2017.

\bibitem{Rantaharju:2019nmh}
Jarno Rantaharju, Claudio Pica, and Francesco Sannino.
\newblock {Walking Dynamics Guaranteed}.
\newblock 10 2019.

\bibitem{Foadi:2012bb}
Roshan Foadi, Mads~T. Frandsen, and Francesco Sannino.
\newblock {125 GeV Higgs boson from a not so light technicolor scalar}.
\newblock {\em Phys. Rev. D}, 87(9):095001, 2013.

\bibitem{Sannino:2004qp}
Francesco Sannino and Kimmo Tuominen.
\newblock {Orientifold theory dynamics and symmetry breaking}.
\newblock {\em Phys. Rev. D}, 71:051901, 2005.

\bibitem{Dietrich:2006cm}
Dennis~D. Dietrich and Francesco Sannino.
\newblock {Conformal window of SU(N) gauge theories with fermions in higher
  dimensional representations}.
\newblock {\em Phys. Rev. D}, 75:085018, 2007.

\bibitem{Dietrich:2005jn}
Dennis~D. Dietrich, Francesco Sannino, and Kimmo Tuominen.
\newblock {Light composite Higgs from higher representations versus electroweak
  precision measurements: Predictions for CERN LHC}.
\newblock {\em Phys. Rev. D}, 72:055001, 2005.

\bibitem{Cacciapaglia:2020kgq}
Giacomo Cacciapaglia, Claudio Pica, and Francesco Sannino.
\newblock {Fundamental Composite Dynamics: A Review}.
\newblock {\em Phys. Rept.}, 877:1--70, 2020.

\bibitem{palatini1919}
A.~Palatini.
\newblock {Deduzione invariantiva delle equazioni gravitazionali dal principio
  di Hamilton}.
\newblock {\em Rend. Circ. Mat. Palermo}, 43:203--212.

\bibitem{RovelliPalatini}
Carlo Rovelli and Francesca Vidotto.
\newblock {\em {Covariant Loop Quantum Gravity}: {An Elementary Introduction to
  Quantum Gravity and Spinfoam Theory}}.
\newblock Cambridge Monographs on Mathematical Physics. Cambridge University
  Press, 11 2014.
\newblock See Eq.(3.30) p.64.

\bibitem{Holst:1995pc}
Soren Holst.
\newblock {Barbero's Hamiltonian derived from a generalized Hilbert-Palatini
  action}.
\newblock {\em Phys. Rev. D}, 53:5966--5969, 1996.

\bibitem{Perez:2005pm}
Alejandro Perez and Carlo Rovelli.
\newblock {Physical effects of the Immirzi parameter}.
\newblock {\em Phys. Rev. D}, 73:044013, 2006.

\bibitem{Rovelli4fermion}
Carlo Rovelli and Francesca Vidotto.
\newblock {\em {Covariant Loop Quantum Gravity}: {An Elementary Introduction to
  Quantum Gravity and Spinfoam Theory}}.
\newblock Cambridge Monographs on Mathematical Physics. Cambridge University
  Press, 11 2014.
\newblock See Eq.(3.32)-(3.34) p.65.

\bibitem{Zubkov:2010sx}
M.~A. Zubkov.
\newblock {Torsion instead of Technicolor}.
\newblock {\em Mod. Phys. Lett. A}, 25:2885--2898, 2010.

\bibitem{Belyaev:1998ax}
A.~S. Belyaev and I.~L. Shapiro.
\newblock {Torsion action and its possible observables}.
\newblock {\em Nucl. Phys. B}, 543:20--46, 1999.

\bibitem{Shapiro:1994vs}
I.~L. Shapiro.
\newblock {The Interaction of quantized matter fields with torsion and metric:
  NJL model}.
\newblock {\em Mod. Phys. Lett. A}, 9:729--733, 1994.

\bibitem{Shapiro:2001rz}
I.~L. Shapiro.
\newblock {Physical aspects of the space-time torsion}.
\newblock {\em Phys. Rept.}, 357:113, 2002.

\bibitem{Rovelli:2014ssa}
Carlo Rovelli and Francesca Vidotto.
\newblock {\em {Covariant Loop Quantum Gravity}: {An Elementary Introduction to
  Quantum Gravity and Spinfoam Theory}}.
\newblock Cambridge Monographs on Mathematical Physics. Cambridge University
  Press, 11 2014.
\newblock See Eq.(3.20) p.63.

\bibitem{Rovelli04}
Carlo Rovelli.
\newblock {\em {\it\,``Quantum gravity``}}.
\newblock Cambridge Monographs on Mathematical Physics. Univ. Pr., Cambridge,
  UK, 2004.
\newblock See Eq.(2.33) p.38.

\bibitem{Zubkov:2014qla}
M.~A. Zubkov.
\newblock {Schwinger\textendash{}Dyson equation and NJL approximation in
  massive gauge theory with fermions}.
\newblock {\em Annals Phys.}, 354:72--88, 2015.

\bibitem{Miransky_1}
V.~A. Miransky.
\newblock {\em {Dynamical symmetry breaking in quantum field theories}}.
\newblock 1994.
\newblock See \S8.4, specifically Eq.(8.49) therein.

\end{thebibliography}
%-------------------------------------------------------------------------------------------------------------------------------------------------------%
\end{document}